\documentclass[aps,prd,twocolumn,nofootinbib,superscriptaddress]{revtex4-1} 
\usepackage{fullpage, amsmath, xparse, amssymb, mathtools, graphicx, braket,xcolor,verbatim}
\usepackage{natbib}
\usepackage[colorlinks=true,linkcolor=blue,citecolor=blue]{hyperref}

\usepackage{multirow}
\usepackage{cancel}

\def\gsim{ \lower .75ex \hbox{$\sim$} \llap{\raise .27ex \hbox{$>$}} }
\def\lsim{ \lower .75ex \hbox{$\sim$} \llap{\raise .27ex \hbox{$<$}} }

\begin{document}
\title{Heating and scattering of stellar distributions by ultralight dark matter}

\author{Andrew Eberhardt}
\thanks{Kavli IPMU Fellow}
\email{andrew.eberhardt@ipmu.jp}
\affiliation{Kavli Institute for the Physics and Mathematics of the Universe (WPI), UTIAS, The University of Tokyo, Chiba 277-8583, Japan}
\author{Mateja Gosenca}
\email{mateja.gosenca@univie.ac.at}
\affiliation{Faculty of Physics, University of Vienna,
  Boltzmanngasse 5, 1090 Vienna, Austria
}
\author{Lam Hui}
\email{lh399@columbia.edu}
\affiliation{Physics Department and Center for Theoretical Physics,
  Columbia University, New York, NY 10027, USA}


\begin{abstract}
Due to wave interference, an ultralight light dark matter halo
has stochastic, granular substructures which can scatter stars,
leading to the heating of stellar distributions.
Studies of this phenomenon have placed lower bounds on
the ultralight dark matter mass.
In this paper we investigate a number of relevant systematic effects, including:
(1) the heating by the central soliton, (2) the self-gravity of the stars,
(3) the suppression of heating in a tidally stripped halo, and
(4) the tidal field suppression of heating when the stellar cluster
is much smaller than the de Broglie wavelength.
The first three effects are quantified by studying the dynamics
of stellar particles in Schr\"odinger-Poisson simulations of
ultralight dark matter halos, while the last effect is studied using
analytic approximations.
\end{abstract}

\maketitle

\section{Introduction}

The current standard model of cosmology, $\Lambda$CDM, has proved an extremely predictive theory, particularly on the large cosmic scales. An essential component of this model is dark matter. However, despite a global effort, the specific dark matter particle has yet to be identified.
Basic properties, such as the dark matter particle's mass, are unknown.
In the vast spectrum of allowed masses, $30$ eV turns out to an interesting scale, below which the de Broglie wavelength $\lambda_\mathrm{dB} = 2 \pi \hbar/m \sigma$ (with $m$ being the particle mass and $\sigma$ being the velocity dispersion) exceeds the typical interparticle separation in a Milky-way like environment \cite{Tremaine:1979we,Hui:2021tkt}.
In this regime, dark matter is necessarily bosonic, and manifests wavelike behavior, resulting in an
array of interesting phenomena. In particular, at the ultralight end of
$m \lesssim 10^{-19} \, \mathrm{eV}$, the wave phenomena have interesting astrophysical implications. 

Historically, much attention has been paid to the mass scale $m \sim 10^{-22}$ eV, advocated by \cite{Hu2000} as fuzzy dark matter to address certain small scale structure problems of conventional
cold dark matter. From a fundamental physics point of view, axions or axion-like-particles make for interesting ultralight dark matter candidates
\cite{Svrcek:2006yi,Arvanitaki_2010,Hui_2017,sheridan2024}, with possible masses spanning a wide range. It is thus interesting to investigate what astrophysical bounds can be put on the mass of ultralight dark matter.

Existing bounds in the literature include: the subhalo mass function ($> 3\times 10^{-21}\, \mathrm{eV}$) \cite{Nadler_2021, Schutz2020},
ultra-faint dwarf half-light radii ($> 3 \times 10^{-19}\, \mathrm{eV}$ \cite{Dalal2022, Marsh:2018zyw}) or cores~\cite{Hayashi:2021xxu}, 
galactic density profiles ($> 10^{-20}\, \mathrm{eV}$ \cite{Bar_2018, Bar_2022}),
satellite masses ($>6 \times 10^{-22}\, \mathrm{eV}$ \cite{Safarzadeh_2020}), 
Lyman-alpha forest ($>2 \times 10^{-20}\, \mathrm{eV}$ \cite{Rogers_2021}), 
and strong lensing ($>4 \times 10^{-21}\, \mathrm{eV}$ \cite{Powell:2023jns}).
On the other hand, it's worth pointing out there are works finding data support $m$ as low as $\sim 10^{-22}$ eV \cite{Broadhurst:2024ggk, Palencia:2025wjw}.
For recent reviews, see \cite{Eberhardt2025Review, Hui_2017, Ferreira_2021,Hui:2021tkt,NIEMEYER2020103787}. There have also been
studies of extensions of the basic model including higher spin \cite{Amin2022, Amin2023}, multiple dark matter fields \cite{Tellez-Tovar2021, Street2022, Guo2021, Luu2020, Davoudiasl2020, Huang2022, Glennon2023, Vogt2022, Chen2023, Gosenca2023,Mirasola2024,Luu2023,luu2024}, self interacting models \cite{Glennon2022,Glennon2023b,Mirasola2024}, and mixed dark matter scenarios.

In this paper, we wish to explore in detail the heating of stellar distributions by their
interaction with halo substructures originating from wave interference.
This interaction has been studied in
analytic computations \cite{Hui_2017,2019ApJ...871...28B} and numerical simulations \cite{DuttaChowdhury2023},
as well as in the context of the Eri II \cite{Marsh:2018zyw}, the milky-way disc \cite{Church2019}, Segue I and II \cite{Dalal2022}, Fornax II dwarf galaxies \cite{teodori2025}, and stellar streams
\cite{Amorisco:2018dcn,Dalal:2020mjw}.
There are a number of relevant systematic effects we investigate further.

(1) While the stochastic heating by granules or quasi-particles is well understood, the heating by soliton random walk and oscillation is less well studied. We fill this gap by performing a series of numerical simulations.
We confirm that the quasi-particle approximation works well in predicting the heating rate of stellar distributions, as long as soliton heating is negligible. When the stars of interest are centrally located,
and the size of the stellar cluster is of the order of the soliton radius or less, the heating by soliton has
to be accounted for and treated differently, and we provide an analytic expression to capture the effect.

(2) In much of the prior work, the self-gravity of stars is assumed unimportant compared to the gravitational force from dark matter. We study cases where this assumption is not valid. We find that when the stellar mass density is comparable to the dark matter density or larger, the half-light radius growth rate is significantly reduced. It's worth emphasizing that a stellar system that is dark matter dominated today could be star dominated in the past. 

(3) As pointed out by \cite{Schive2020,Li:2020ryg,yang2025}, heating is suppressed in a
tidally stripped (sub-) halo, for stars located in the soliton. We find that the heating rate
is not significantly impacted if the degree of tidal stripping is mild. But when the tidal stripping
removes a significant portion of the halo outside the soliton, the heating rate is substantially reduced.

(4) When a stellar cluster has a size much smaller than the de Broglie wavelength, one expects the differential gravitational force across the cluster to be reduced. We study this tidal field suppression effect (to be distinguished from the tidal stripping effect above) using analytic approximations.

How might these systematic effects affect the existing heating constraints on ultralight dark matter? Our goal in this paper is to lay some of the groundwork for addressing this question.
Further work needs to be done though, before we can provide a reliable assessment.
First, the efficiency of tidal stripping for an ultralight dark matter subhalo needs to be better quantified with numerical simulations. As noted by \cite{Hui_2017,Du:2018qor}, the soliton is susceptible to enhanced tidal disruption,
due to quantum pressure. A standard estimate of the tidal radius is not sufficient to capture the effect. More work needs to be done to quantify the tidal disruption of the larger subhalo, including the granules outside the soliton. Recent work by \cite{yang2025} examined this for a satellite galaxy like Fornax. Further investigations for systems similar to ultrafaint dwarf galaxies would be helpful.
Second, some of the effects mentioned above are most significant when the radius of the stellar cluster is much smaller than the de Broglie wavelength (or soliton radius). This is particularly relevant for the low end of the dark matter mass spectrum, with its correspondingly large de Broglie wavelength. 
It's also a regime challenging to simulate because of the wide separation of scales. We hope to address both issues in future work.


The paper is organized as follows: Section \ref{sec:background} contains a background discussion on ultralight dark matter and
the construction of halos to initialize Schr\"odinger-Poisson simulations.
In Section \ref{sec:heating}, we review the basics of heating by quasi-particles, and discuss analytic approximations that appear to work well when soliton heating is non-negligible. 
Section \ref{sec:simulations} contains a discussion of our simulation method, including the set-up of initial conditions and the subsequent evolution by the Schr\"odinger-Poisson system of equations.
In Section \ref{sec:results} we go over in detail the results of our many numerical experiments.
We conclude in \ref{sec:conclusion}.

In Appendix \ref{app:PlummerICs} we explain how the initial stellar distributions are sampled. In Appendix \ref{app:segueii} we verify that our analytic approximation reproduces an example given in \cite{Dalal2022}, under the same set of assumptions.
In Appendix \ref{app:smallR} we investigate what happens when the stellar cluster has an initial size much smaller than the de Broglie wavelength. We use analytic approximations to estimate the effects of tidal field suppression and the stellar self-gravity, including a discussion of Segue 1.

\section{Background} \label{sec:background}

\subsection{Schr\"odinger-Poisson}

\begin{figure*}[!ht]
	\includegraphics[width = .97\textwidth]{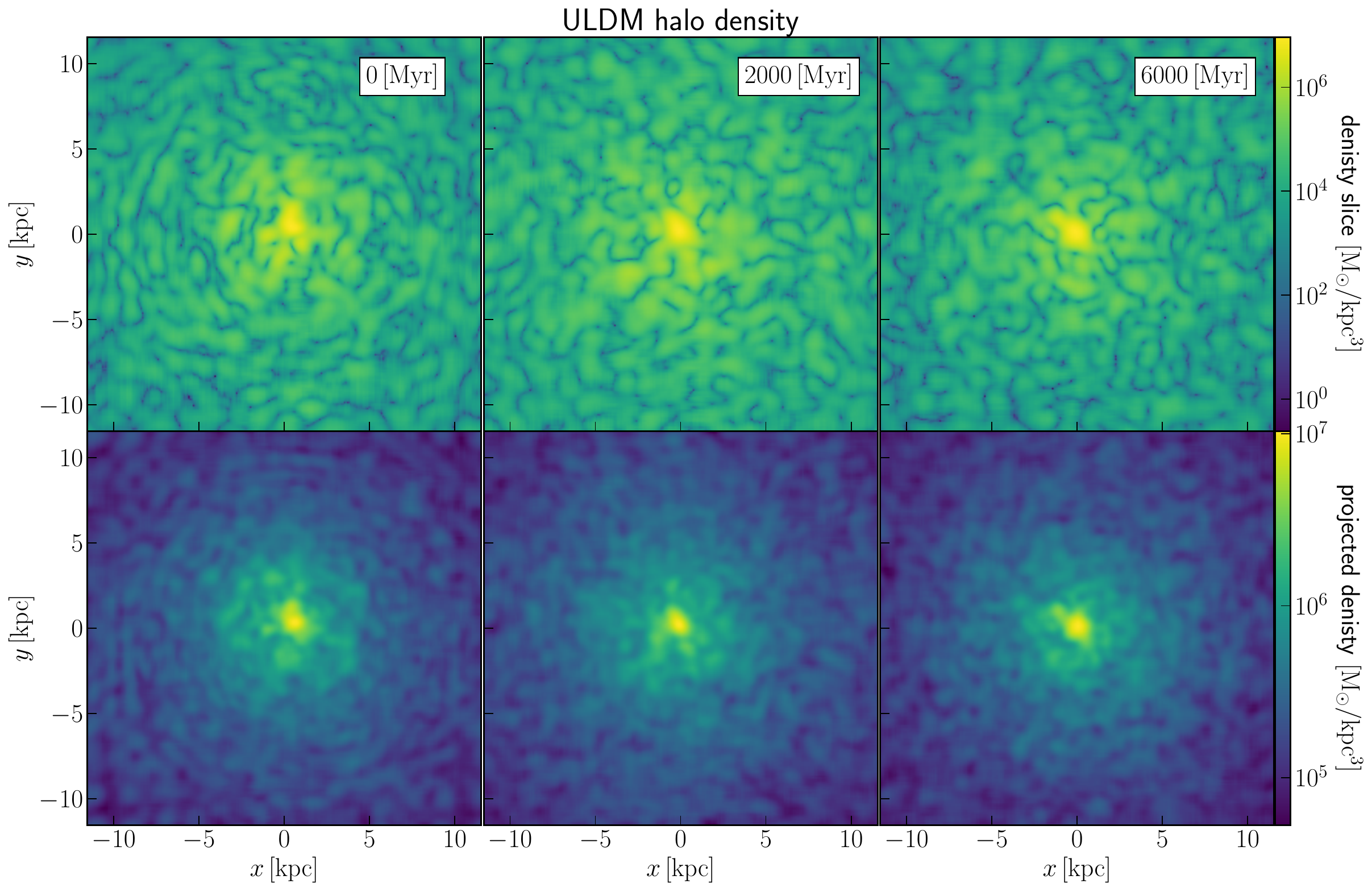}
	\caption{ Evolution of an ultralight dark matter field halo. Initial conditions are produced using the eigenmode method and then simulated using the full nonlinear Schr\"odinger-Poisson equations. Columns show three simulation snapshots. The top and bottom rows show density slices and projections, respectively. The halo is relatively stable throughout its evolution, although the soliton moves randomly around the center. Here the ULDM mass is $m_{22} = 5$, the scale radius is $R_s = 2 \, \mathrm{kpc}$, and scale density is $\rho_0 = 1.89 \times 10^{6} \, \mathrm{M_\odot / kpc^3}$. Slight azimuthal alignments visible in the first snapshot are due to the finite number of eigenfunctions used at initialization. They relax after a few time steps of the simulation.}
	\label{fig:UDLM_halo_sim}
\end{figure*}

When simulating ultralight dark matter we often approximate the field as classical and non-relativistic. The motivation for these approximations has been studied in great detail in a large body of work \cite{Eberhardt2023}. In the non-relativistic and classical limits ultralight dark matter is usually modeled using as one or more classical fieldsbeying the Schr\"odinger-Poisson equations given by
\begin{align} \label{eqn:SP}
    i \hbar \, \partial_t \psi(x) &= \left(-\frac{\hbar^2}{2 m } \nabla^2 + m V(x) + m V_{\mathrm{ext}}(x) \right) \,\psi_i(x) \,  \nonumber \\
    \nabla^2 V(x) &= 4 \pi G \, m |\psi(x)|^2 \, ,
\end{align}
where $\psi$ and $m$ are the field and field mass respectively. In principle $V_\mathrm{ext}$ represents any external potential but in this work we will use it to represent the gravitational potential due to cold dark matter. There exists work investigating higher spin and self-interacting fields as well. The square amplitude of the field is interpreted as a density 
\begin{align}
    \rho(x) =  m |\psi(x)|^2 \, .
\end{align}
The classical field description of these systems results in a few interesting features. First, there are interference patterns in the density at the scale of the de Broglie wavelength given by
\begin{align}
    \lambda_\mathrm{dB} = \frac{2 \pi \hbar}{m \, \sigma_\mathrm{DM}} \, .
\end{align}
Here, $\sigma_\mathrm{DM}$ 
is the dark-matter velocity dispersion. This creates a density field made up of ultralight dark matter ``granules". The granular structure is not constant and changes with de Broglie time approximately equal to the crossing time of the granules, i.e.
\begin{align}
    \tau_\mathrm{dB} = \frac{\lambda_\mathrm{dB}}{\sigma_\mathrm{DM}} \, .
\end{align}
Additionally, each ultralight dark matter halo has at its core a ``soliton" which provides a cored rather than cuspy profile. The soliton is the ground-states solution to equations \eqref{eqn:SP} and generally stochastically ``walks" around the center of the halo. Figure \ref{fig:UDLM_halo_sim} shows the dynamics of an ULDM halo near a steady-state.

\subsection{ULDM halo construction} \label{subsec:HaloConstruct}

\begin{figure}[!ht]
	\includegraphics[width = .44\textwidth]{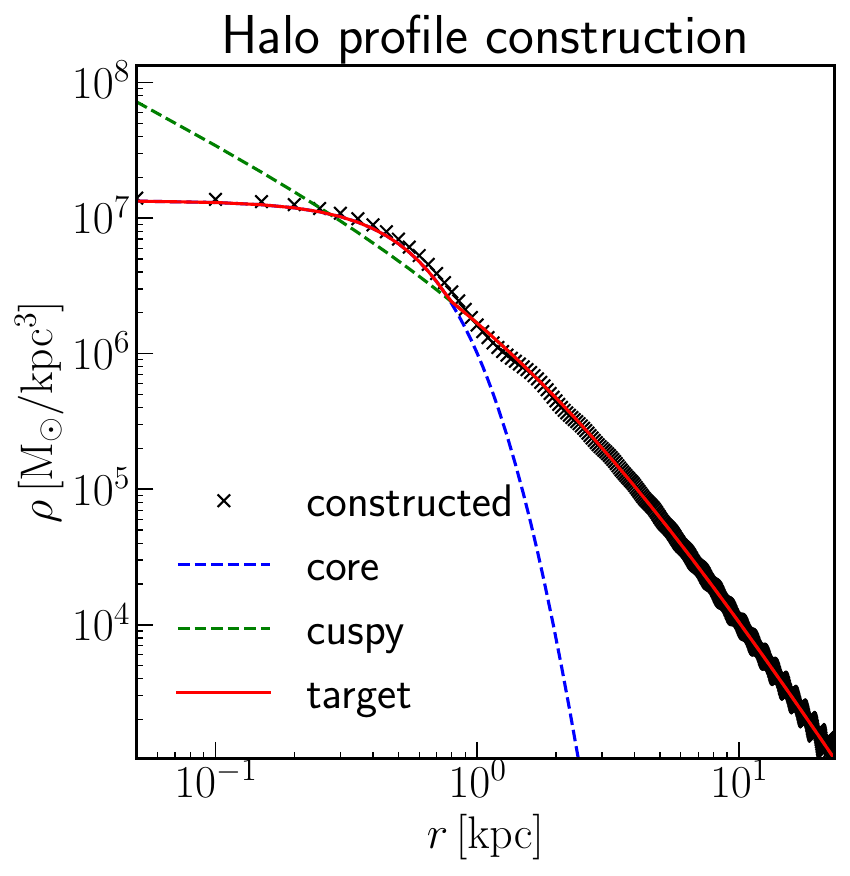}
	\caption{ Halo density profiles used in halo generation. The target profile, red, is given as a combination of a cuspy NFW profile, green, beyond the core radius and a core profile, blue, inside the core radius. The constructed profile, black, is composed of the best fit sum of weighted eigenmodes of the Hamiltonian given by the potential from the target density profile. Here the ULDM mass is $ m_{22} = 5$, the halo profile scale radius is $R_s = 2 \, \mathrm{kpc}$, and scale density is $\rho_0 = 1.89 \times 10^{6} \, \mathrm{M_\odot / kpc^3}$. }
	\label{fig:profiles}
\end{figure}

In order to explore stellar configurations in individual halos efficiently, mock halos are constructed using the eigenmode method used in several previous works \cite{Yavetz2022, Zagorac2022}. We use this method to construct initial conditions before evolving the system using the nonlinear Schr\"odinger-Poisson equations.
We first choose a ``target" density profile, $\rho_t$, representing the near-stable spherically-symmetric solution for a given set of halo parameters and ultralight field mass and then construct the ultralight dark-matter field as a sum of the eigenmodes associated with the Hamiltonian given by the target profile and equations \eqref{eqn:SP}. The target profile is a cored NFW profile given by
\begin{align} \label{eqn:coredNFW}
    \rho_{t}(r) = \left\{
\begin{array}{ll}
      \frac{(1.9\times 10^{7}) \, m_{22}^{-2} r_c^{-4}}{(1+0.091 \, (r/r_c)^2)^8}   \,,& r < r_t \\
      \frac{\rho_0}{\frac{r}{R_s} \left( 1 + \frac{r}{R_s} \right)^2} \,,& r \geq r_t \\
\end{array} 
\right. \, ,
\end{align} 
in units of $\mathrm{M_\odot / kpc^3}$. Here, $\rho_0$ is the scale density, $R_s$ the scale radius, $r_c$ the core radius, and the field mass is $m = m_{22} \times 10^{-22} \, \mathrm{eV}$. $r_t$ is the transition radius, it is $\sim r_c$, and is chosen such that the density is continuous. Figure \ref{fig:profiles} contains a comparison of these profiles. 

The density of the target profile enters the Hamiltonian, $H$, via the radial Schr\"odinger-Poisson equations, i.e. 
\begin{align} \label{eqn:}
    &H_l = -\frac{\hbar^2}{2 m } \frac{1}{r^2} \partial_r \left( r^2 \partial_r \right) + m V_{\rm t}(r) + \frac{\hbar^2}{2m} \frac{(l + 1)l}{r^2} \, , \nonumber \\
    &\nabla^2 V_{\rm t}(r) = 4 \pi G \, \rho_{\rm t}(r) \, ,
\end{align}
where $l$ is the angular momentum quantum number and $V_{\rm t}$ is the gravitational potential from the target density profile obtained by solving the above radially-symmetrized Poisson equation for $\rho_{\rm t}$. In this work we will only construct statistically isotropic halos, but eigenmode constructions of halos with triaxiality has been studied previously \cite{Zagorac2022}. We numerically find the eigenvalue-eigenmode pairs associated with this Hamiltonian, i.e. 
\begin{align}
    H_l \, u_n^l = E_n^l \, u_n^l \, .
\end{align}
Here, the functions $u$ are related to the eigenvectors as $u = r\phi$. The eigenvectors are then fitted with weights, $w_j$, such that their sum best approximates the target profile, i.e. 
\begin{align}
    \rho_{\rm t}(r) &\approx \sum_l^{l_{\rm max}} \sum_{m = -l}^{l} \sum_{n}^{n_{\rm max}} w_n \, |\phi_n^l(r)|^2 \, .
\end{align}
We truncate the sum at $n_{\rm max}$, which is associated with the highest-energy eigenfunction that is still included. Our isotropy condition means that the weights and eigenfunctions do not depend on the angular momentum in the $z$ direction, and thus the middle sum simply provides a factor of $(2l + 1)$. Notice that our truncated basis of squared eigenfunctions is neither orthonormal nor complete and so we iteratively optimize to find the best fit values of the weights \footnote{using scipy's optimize module.}. The resulting profile does not exactly match the target profile, but is a very good approximation,  see Figure \ref{fig:profiles}.

The final field is composed of the eigenfunctions summed using the weights and a random phase, i.e.
\begin{align} \label{eqn:eigSum}
    \psi(r, \theta, \varphi) = \sum_l^{l_{\rm max}} \sum_{m = -l}^{l} \sum_{n}^{n_{\rm max}} w_n \, Y^m_l(\theta, \varphi) \, \phi_n^l(r) \, e^{-i \, \alpha_{lmn}} \, .
\end{align}
$\alpha_{lmn}$ is chosen uniformly and randomly from $[0,2 \pi)$. This is then the initial condition of our field.

\section{Heating of stellar dispersions} \label{sec:heating} 

In this Section we discuss the heating of stars in ultralight dark matter halos. To quantify how the size of the halo changes with time, we trace the characteristic radius $R$. 
The transfer of energy causes the half-light radius of an embedded stellar distribution to grow over time. We will split the integration of the half-light radius into four parts.
First, we will discuss a basic introduction to the heating of particles in a ULDM background.
Then in a halo, we split the discussion into the regime in which the heating due to the soliton is important and where it is not. When the soliton is not important we can simply apply the standard quasi-particle approximation which has been widely used to approximate the heating due to ultralight dark matter granules \cite{Gosenca2023, Dalal2022, teodori2025, Marsh:2018zyw, Hui_2017}. Next, when the soliton is important, the heating is described by the average potential fluctuations which is not well described by the standard quasi-particle approximation. Finally, we will discuss the linear approximation at radii small compared to the core. 

\subsection{Basics of stochastic heating by granules}
\begin{figure*}[!ht]
	\includegraphics[width = .97\textwidth]{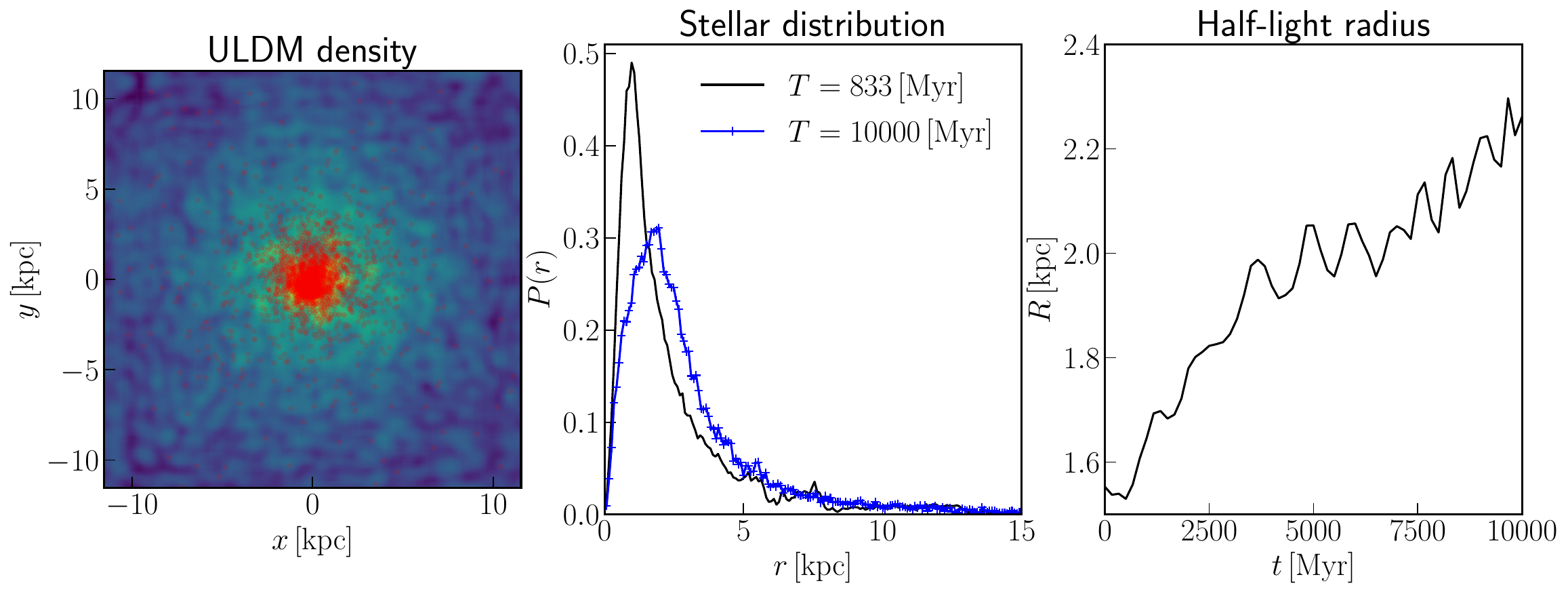}
	\caption{ \textbf{Left:} the log density of an ultralight dark matter halo. The red dots superimposed on the density show the projected position of stellar particles. \textbf{Center:} distribution of stellar particle radii embedded in the halo at two different snapshots. \textbf{Right:} Half-light radius of stellar particles overtime. We can see the half-light radius of the particles increases overtime consistent with the heating from the ultralight dark matter halo. In this simulation $m_{22} = 5$ and $f_\mathrm{ULDM} = 0.5$ ($f_\mathrm{ULDM}$ defined in equation \eqref{eqn:define_f}). }
	\label{fig:heating_specific}
\end{figure*}


Due to interfering waves, the ULDM halos exhibit $\sim\mathcal{O}(1)$ density fluctuations. This granular structure is visible in Figure \ref{fig:UDLM_halo_sim}. Our derivation follows \cite{Hui_2017}. In order to describe stellar heating analytically, these density granules are sometimes approximated as quasiparticles with a radius similar to the de Broglie wavelength, i.e. $r \sim \lambda_\mathrm{dB}$. 
This means that the approximate granule mass $M_{\rm eff}$ is 
\begin{align}
    M_\mathrm{eff} \sim \left( \frac{\pi \hbar}{m \, \sigma_{\rm DM}} \right)^3 \rho
\end{align}
Where $\rho$ and $m$ are the density and mass associated with the ultralight field producing the granule, and $\sigma_{\rm DM}$ is the local dark matter velocity dispersion. These granules provide gravitational kicks to passing stars injecting energy into the stellar distribution.
In the weak deflection limit we can write the change in a star's velocity due to an encounter with a granule of mass $M_{\rm eff}$ as
\begin{align}
    \delta v \sim \frac{G M_\mathrm{eff}}{\lambda_\mathrm{dB} v_{\rm r}} \, ,
\end{align}
where $v_r$ is the typical relative velocity between a given star and a given granule.

A set of $N$ stars with velocities $\{v_i\}$ has the velocity expectation value 
\begin{equation}
    \bar{v} = \frac{1}{N}\sum_{i=0}^Nv_i \: , 
\end{equation}
and the variance of stellar velocities 
\begin{equation}
    \mathrm{var}[v] = \frac{\sum_{i=0}^N(v_i - \bar{v})^2}{N}.
\end{equation}
After this system has encountered $N_\mathrm{enc}$ granules, the expectation value of velocities stays the same, 
\begin{equation}
    \bar{v}_{\rm fin}  = \bar{v}_{\rm ini}, 
\end{equation}
because we assume that gravitational kicks $\delta v$ are normal-distributed and uncorrelated, $\delta v = \mathcal{N} (0, \sigma_{\delta}^2)$, around the mean value of $0$ and with dispersion equal to the typical magnitude of the kicks, $\sigma_{\delta}^2 = \delta v^2$. However, the variance changes by 
\begin{equation}
\begin{split}
    \Delta \mathrm{var}[v] &= \mathrm{var}[v]_{\rm fin} - \mathrm{var} [v]_{\rm ini} \\ 
    & =  \frac{1}{N} \sum_{i=0}^N \sum_{j=0}^{N_\mathrm{enc}} \left( v_{i} +\delta v_j - \bar{v}_\mathrm{fin}\right)^2 -  \left(v_{i} - \bar{v}_{\mathrm{ini}} \right)^2 \\
    & \simeq N_\mathrm{enc} \delta v ^2. 
\end{split}
\end{equation}
For a particle with constant speed, the number of encounters is the total distance traveled, divided by one de Broglie length:  
\begin{equation}
    N_\mathrm{enc} = \frac{v_r t}{\lambda_\mathrm{dB}}.
\end{equation}
Note that $v_r$ may differ from the velocity dispersion of the dark matter if the stars and the dark matter have different velocity distributions.

If we assume the kicks are uncorrelated, we can write the change in the variance as a sum of consecutive kicks as 
\begin{align}
    \partial_t \mathrm{var}[v] \sim 
    \delta v^2 \frac{dN_\mathrm{enc}}{dt} \, .
\end{align}
Where the number of encounters per time is given by
\begin{align}
    \frac{dN_\mathrm{enc}}{dt} \sim \frac{v_{\rm r}}{\lambda_\mathrm{dB}} \, .
\end{align}
The change in the variance can then be written as
\begin{align} \label{eqn:deltaVar}
    \Delta \mathrm{var}[v] &\sim \frac{G^2 M_\mathrm{eff}^2}{\lambda_\mathrm{dB}^3 v_{\rm r}} \Delta t \\
    &= \frac{C \pi^3 G^2 \hbar^3 \rho_\mathrm{ULDM}^2}{m^3 \sigma_{\rm DM}^3 v_{\rm r}} \Delta t \, .
\end{align}
Where $C \sim \mathcal{O}(1)$ 
is some constant which depends on the details of the stellar and dark matter distributions. 

\subsection{Far from the core} \label{sec:farFromCore}

In the limit where the stellar half-light radius is large compared to the de Broglie wavelength and core radius, $R \gg r_c$, the quasi-particle approximation describes the transfer of energy between the halo and stars. This is because in the outer halo the density oscillations that most stars encounter are locally well described by the constant stream of uncorrelated density granules that we assumed in the previous Section. However, because the stars are no longer free but in a distribution we will need to consider how the gravitational kicks affect the energy of the distribution instead of the velocity variance. 

Because the system is virialized and the kinetic energy, like the variance, is a second moment of the particle speeds, we expect the change in energy to be related to the change in the variance for free particles by some constant. Let us say then that the right side of equation \eqref{eqn:deltaVar} sources the change in the energy per particle mass as
\begin{align} \label{eqn:dE_far}
    \partial_t E = \, \frac{C G^2 \pi^3 \hbar^3 \rho_\mathrm{ULDM}^2}{m^3 \sigma^4}  \, .
\end{align}
 The velocity distributions of the dark matter and stars 
 are now related parameters since they are both virialized in the same net potential, $\sigma_\mathrm{DM}(r) = \sigma_*(r) = \sigma(r) \sim \sqrt{G M_\mathrm{enc}(r) / r}$. 

We approximate the distribution using the values of density and velocity evaluated at the half-light radius. To calculate the change in radius let us consider a test particle on a circular orbit at $R$. 
The energy per mass of the particle is the sum of the kinetic and potential parts, $E= T+U$, with $U(r) = -\int_\infty^{r} {\bf F}\cdot d\mathcal{\bf l}$. Therefore, the energy per mass is\footnote{Here, we integrate from $0$ rather than the usual convention of integrating from the infinity. The energy definition above differs from the conventional definition by a constant.}
\begin{equation} \label{eqn:total_energy}
\begin{split}
E(R) &= \frac{v(R)^2}{2} + \Phi (R) \\
    &= \frac{G \, M_\mathrm{enc}(R)}{2R} + \int_0^{R } \frac{G \, M_\mathrm{enc}(r) }{r^2} dr.
\end{split}
\end{equation}
Let us consider a small change in the energy $\delta E$ and look at the corresponding change in the radius: 
\begin{equation} \label{eqn:energy_change_halo}
\begin{split}
    \delta E &= \left.\frac{\partial E}{\partial r}\right|_{r=R} \delta R \\ 
    &= \Bigg(-\frac{G \, M_\mathrm{enc}(R)}{2R^2} +\frac{G \, \partial_{r} \left.M_\mathrm{enc}(r)\right|_{r=R}}{2R} \\
    &+  \left.\partial_{R} \int_0^{R} \frac{G \, M_\mathrm{enc}(r') }{r'^2} dr'\right|_{r=R} \Bigg) \delta R\\
    & = \frac{1}{2} \frac{G M_\mathrm{enc}(R)}{R^2} \delta R + \frac{G \, \partial_{r} \left.M_\mathrm{enc}(r)\right|_{r=R}}{2R} \delta R.
\end{split}  
\end{equation}
Here, we use the Leibniz integral rule. If we assume that the enclosed mass at $r=R$ is varying slowly compared to its average variation inside the half-light radius, $\partial_r M_\mathrm{enc}|_{r=R} \ll M_\mathrm{enc}/R$, then the second term vanishes and we have $\delta E = \frac{1}{2} \frac{G M_\mathrm{enc}(R)}{R^2} \delta R$. Whether this assumption is valid depends on the density profile of the halo. For example, in a purely NFW profile it holds in the outer region of the halo. In the inner region, $\partial_r M_\mathrm{enc}|_{r=R} = 2 M_\mathrm{enc}/R$ and we get $\delta E = \frac{3}{2} \frac{G M_\mathrm{enc}(R)}{R^2} \delta R$. To account for both these limits and also include other scenarios, such as a profile with a central soliton, we can write
\begin{equation}
     \delta E = \frac{\alpha}{2} \frac{G M_\mathrm{enc}(R)}{R^2} \delta R
\end{equation}
where $\alpha = 1 + R\partial_r M_\mathrm{enc}(r)|_{r=R}/ M_\mathrm{enc}$.
The change in energy as a function of a small change in $R$ is 
\begin{align}
    \delta E = \alpha\frac{ F \, \delta R}{2} 
\end{align}
Where $F = -\vec F \cdot \hat r = G M_\mathrm{enc} / R^2$ is the central force on the particle. 
Solving for $\delta R$ we then find 
\begin{align} \label{eqn:deltaR_skirt}
    \delta R = \frac{2 \delta E}{\alpha F} \, .
\end{align}
We will approximate the evolution of the half-light radius using the values of the central force and energy evaluated at the half-light radius. The evolution of the half-light radius is therefore given by the following integration 
\begin{align} \label{eqn:Rhalf_predict}
    R(t) \approx \int_0^t  \frac{2 \, \partial_t E(t'; R(t') )}{ \alpha F(t'; R(t') )} dt' \, .
\end{align}
Where 
\begin{align} \label{eqn:dr_predict_largeR}
    \partial_t E(t'; R(t') ) &= \frac{C \pi^3 \, \hbar^3 \rho_\mathrm{ULDM}^2(R(t')) }{m^3 \sigma^4} \, , \\
    F(t'; R(t') ) &= \frac{\alpha G M_\mathrm{enc}(R(t')) }{R^2(t')} \, . \label{eqn:force} 
\end{align}
where $C \sim \mathcal{O}(1)$. $M_\mathrm{enc}(R(t'))$ is the total mass enclosed at $R$ at time $t'$. 

We have derived the heating rate in this section in the massless star limit, i.e. $\rho_* \ll \rho_\mathrm{DM}$. In the limit where the stellar self-potential is important, equation \eqref{eqn:energy_change_halo} needs to be amended to include the self-energy of the stellar distribution. This may significantly complicate equation \eqref{eqn:deltaR_skirt}. 

\subsection{Near the soliton}

The heating in and around the central soliton is not well described by the quasi-particle approximation which we had found in the previous section to be accurate in the outer halo. This idea is supported by the fact that the assumption of an uncorrelated head wind of density granules breaks down near the soliton. Instead there is a single overdensity which oscillates and randomly walks about the base of the potential. In this section we will discuss how the heating due to the soliton can be quantified. 

The system is near a steady state but the oscillation and random walk of the central soliton creates some temporal variation of the potential. Let us consider the potential fluctuations around the spherically averaged potential as
\begin{align} 
    &\Phi_\mathrm{rms}(r) = \label{eqn:phi_rms_define} \\
    &\sqrt{  
    \int_0^T \frac{dt}{T}\int_\Omega \, \frac{d \Omega}{4 \pi} \, \left( \Phi^\mathrm{ULDM}(r,\Omega,t) - \Phi^\mathrm{ULDM}( r,t) \right)^2 } \, . \nonumber 
\end{align}
Where $\Omega$ are the angular coordinates, and $\Phi(r,\Omega,t)$ is the specific value of the potential at a position in space $(r, \Omega)$, $T$ is the total simulation time. $\Phi(r,t)$ is the spherically averaged potential. We will say that the potential fluctuates with root-mean-square amplitude $\Phi_\mathrm{rms}$ at temporal intervals of $\tau_\mathrm{dB}$. Because the change in energy per mass is given by these fluctuations, the change in energy for a particle at an initial radius $r$ 
after a time $\delta t$ is given by
\begin{align} \label{eqn:delta_E_soliton}
    \delta E(r) \propto \, \Phi_\mathrm{rms}(r) \frac{\delta t}{\tau_\mathrm{dB}} \, .
\end{align}
The rms potential fluctuations is given by
\begin{align}\label{eqn:phi_rms}
   \Phi_\mathrm{rms}(r) \propto \frac{G M_\mathrm{enc}^\mathrm{ULDM} }{r} \frac{\lambda_\mathrm{dB}}{r} \, .
\end{align}
This equation is a fit to the simulated data, see Figure \ref{fig:phi_fluctuations}. With this, we can write the change in energy as 
\begin{align} \label{eqn:dE_nearCore}
    \partial_t E = \frac{b \,G M_\mathrm{enc}^\mathrm{ULDM} }{r^2} \frac{\lambda_\mathrm{dB}}{  \tau_\mathrm{dB}} \, ,
\end{align}
where $b$ is some constant that we find to consistently be around $\sim 1/60$ in our simulations (see Section \ref{sec:solitonHeating}).

Let us approximate the energy of the particle distribution as being well described by the energy of a circular orbit at the half-light radius $E \approx E(R)$. We can estimate a potential energy for the particles at the half-light radius by
\begin{align}
    U(R) = \int_0^R \frac{G M_\mathrm{enc}^\mathrm{DM}(r') }{r'^2} dr' \, .
\end{align}
The kinetic energy at $R$ can be approximated by
\begin{align}
    T(R) = \frac{1}{2} \frac{G M_\mathrm{enc}^\mathrm{DM}(R)}{R} \, , 
\end{align}
and the total energy is then simply the sum
\begin{align} \label{eqn:totalEnergy}
    E(R) = U(R) + T(R) + E_*(R)\, .
\end{align}
The main difference between this equation and equation \eqref{eqn:total_energy} is that here $U(R)$ and $T(R)$ contain only the dark matter component, with the stellar self-energy $E_*(R)$ approximated as $ \sim -\frac{GM_*}{R}$.
 
For particles distributed in a halo this function will be invertible, let us therefore define the following function $g$ as
\begin{align} \label{eqn:f_define}
    R \equiv g(E) \, .
\end{align}
We note that in $f$ does not always have a closed form solution. The half-light radius as a function of time is given by the following integration 
\begin{align} \label{eqn:Rhalf_soliton}
    R(t) = g \left( E_i + \int_0^t b \frac{G M_\mathrm{enc}^\mathrm{ULDM}(R (t')) \lambda_\mathrm{dB}}{R^2 (t')} \frac{dt'}{\tau_\mathrm{dB}} \right) \, .
\end{align}

\subsection{Inside the core} \label{sec:insideCore}

Here we consider radii $R \lesssim r_c$. We will discuss this limit for massless particles (in appendix \ref{app:smallR} we discuss how this limit changes for massive particles and for $R \ll \lambda_\mathrm{dB}$). We point out however, in this work we do not simulate the regime where $R \ll \lambda_\mathrm{dB}$. In this limit the density profile is approximately constant, $\rho_c$, i.e., 
\begin{align}
    \rho(r) = \rho_c + \mathcal{O}(r^2/r_c^2) \, .
\end{align}
The energy of the system is given as 
\begin{align}
    E(R) &\approx \frac{GM_\mathrm{enc}(R)}{2 R} + \int_0^{R} \frac{GM_\mathrm{enc}(r)}{r^2} dr \\
    &= \frac{4 \pi G \rho_c R^2}{3} 
\end{align}
Where we have again approximated the distribution using only the values of energy evaluated at the half-light radius.

In this regime we can rewrite the potential fluctuations 
using equation \eqref{eqn:phi_rms} as
\begin{align}
    \Phi_\mathrm{rms} = \frac{b G M_\mathrm{enc} \lambda_\mathrm{dB} }{r^2} = \frac{4 \pi}{3} b \rho_c G r \lambda_\mathrm{dB} \, .
\end{align} 
The heating rate in this regime is then given as
\begin{align} \label{eqn:dE_near}
    \partial_t E = \Phi_\mathrm{rms} \frac{1}{ \tau_\mathrm{dB}} = \frac{4 \pi b \rho_c r G \lambda_\mathrm{dB}}{3 \tau_\mathrm{dB}}  \, .
\end{align}
We obtain the last equality using the fact that $\lambda_\mathrm{dB} = \sigma \, \tau_\mathrm{dB}$. So the heating does not explicitly depend on the field mass.

The change in the half-light radius can then be calculated using equation \eqref{eqn:Rhalf_predict}, in this limit this simplifies to be
\begin{align}
    \partial_t R = \frac{b}{2} \sigma  \, . \label{eqn:linearApprox}
\end{align}

\section{Simulations} \label{sec:simulations}

\subsection{Initial conditions} \label{sec:simICs}

Here we discuss initial condition generation for the simulations presented in this work. We will discuss the initial conditions for both ultralight dark matter field and particles. We will produce initial conditions using summations of eigenmodes and then simulate these systems using the full nonlinear Schr\"odinger-Poisson equations. 

\subsubsection{Ultralight dark matter initial conditions} \label{subsec:ULDM_ICs}

\begin{figure*}[!ht]
	\includegraphics[width = .99\textwidth]{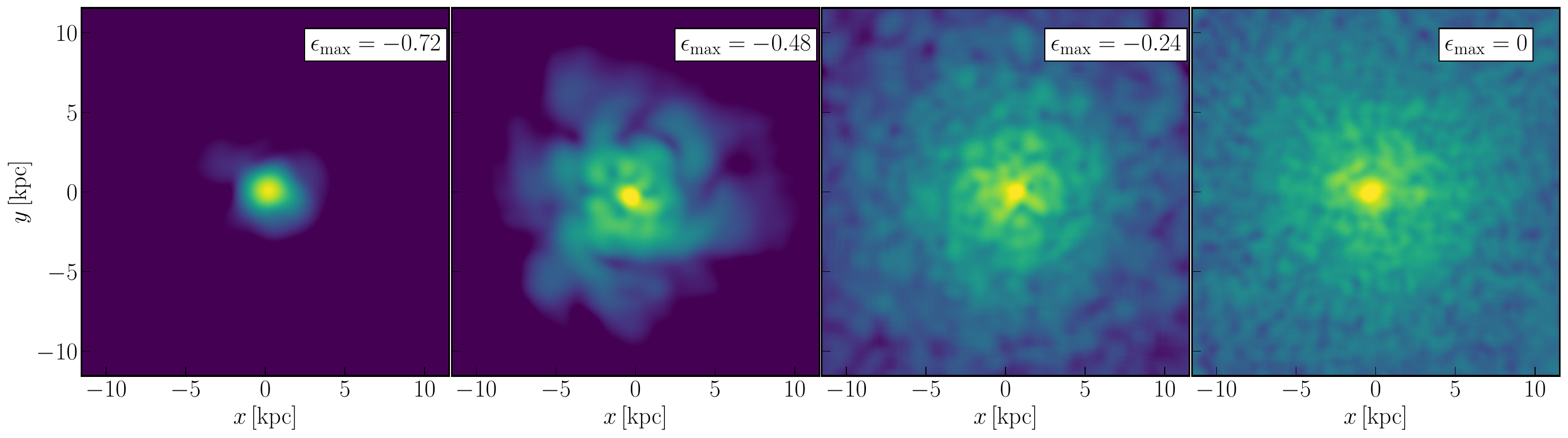}
	\caption{ Halos constructed with different values of $\epsilon_\mathrm{max}$. \textbf{Left:} a halo containing only the central soliton and a few of the next lowest eigenmodes. The granularity is the outer halo is entirely absent. \textbf{Center:} a halo containing the soliton and low energy modes. The granularity in the density has begun to develop, particularly in the vicinity of the core. \textbf{Right:} the benchmark halo. The granular structure is present throughout the halo. $\epsilon_\mathrm{max} \equiv E_\mathrm{max} / |E_0|$ is the maximum energy as a fraction of the ground state energy. Eigenmodes with energy exceeding $E_\mathrm{max}$ are removed in the halo construction.}
	\label{fig:strippedHalos}
\end{figure*}

\textbf{Plane-wave box.}
In these simulations we construct an evolving ultralight dark matter field as a sum of planewaves in a periodic box. It has been demonstrated that this approximates the behavior of ultralight dark matter density for localized regions of ultralight dark matter halos that are small compared to the dynamic radius of the halos \cite{Eberhardt2024,Eberhardt2025}. We will use these simulations to study how heating scales with changes in physical parameters. 

We initialize the field as a sum of planewaves as follows
\begin{align}
    \psi(\vec x) = \sqrt{\rho_0} \sum_i e^{-im\, x \, \vec v_i / \hbar + i \phi_i} \, .
\end{align}
$\phi_i$ is a random uniform drawn from $[0,2 \pi)$. Where $\vec v_i$ are drawn from an isotropic Maxwell Boltzmann distribution, i.e.,
\begin{align}
    f(v) = \sqrt{\frac{2}{\pi}} \frac{v^2}{\sigma^2} e^{-v^2 / 2 \sigma^2} \, , \label{eqn:maxwellian}
\end{align}
with parameter $\sigma$ describing the velocity dispersion. We note that due to the periodic boundary conditions we choose $v_i$ such that the field identifies at the boundaries of the box, i.e., $\psi(-L/2) = \psi(L/2)$. The field is normalized so that the density is $\rho_0$. 


\textbf{Halos.} We initialize an ultralight dark matter field according to the eigenvector summing scheme described in Section \ref{subsec:HaloConstruct}. The halo produced by this method is then simulated using the Schr\"odinger-Poisson equations in isolation for $\sim 1 \, \mathrm{Gyr}$ to remove numerical artifacts resulting from the eigenmode method. Being a sum of eigenfunctions of the radially averaged profile, the halo is in an near-steady state, see Figure \ref{fig:UDLM_halo_sim}.

To model tidally stripped halos, we modify our eigenmode summation to throw out eigenvalues above some energy threshold. This results in halos with smaller mass and fewer granules around the soliton.
The maximum eigenenergy used in the halo construction is a free parameter and we introduce the cutoff energy $\epsilon_\mathrm{max} \equiv E_\mathrm{max} / |E_0|$, where $E_0$ is the energy of the ground state. Some example halos created using this method are plotted in Figure \ref{fig:strippedHalos}.

We also explore ``mixed'' dark matter halos in which ultralight dark matter is just a fraction, $f_\mathrm{ULDM}$, of the total dark matter, 

\begin{equation} \label{eqn:define_f}
    f_\mathrm{ULDM} = \frac{\rho_\mathrm{ULDM}}{\rho_\mathrm{ULDM}+\rho_\mathrm{CDM}}.
\end{equation}
In this case, cold dark matter density is represented as a spherically-symmetric NFW profile whose center overlaps with the center of the simulation box.

Because the ultralight dark matter halo is a sum of eigenvectors which are approximately eigenvectors of an NFW potential (with the exception of those near the core) changing the fraction of ultralight dark matter does not change the shape of ultralight dark matter profile much. 
The gravitational potential due to dark matter is therefore given by
\begin{align}
    V_{\rm tot} =  V_{\rm ULDM} + V_{\rm NFW} \, .
\end{align}
Where $V_{\rm NFW}$ has the same scale radius as the profile used in the eigenmode construction of the ultralight dark matter halo. The idea that cold dark matter should still form a cuspy profile when in mixed ultralight-cold settings is supported by simulations. 

\subsubsection{Particle initial conditions}

\textbf{Stellar particles with Plummer distribution}
In these simulations, the stellar particles are treated as massive point particles. Positions and velocities are chosen using an adapted version of a Plummer sphere scheme \cite{Aatseth1974}. 
The polytropic model with $n=5$, also known as the Plummer sphere, is characterized by the density profile 
\begin{equation}\label{eq:Plummer_rho}
    \rho(r) = \frac{3M_{*}}{4\pi R^3}\left( 1 + \frac{r^2}{R^2}\right)^{-5/2}
\end{equation}
for some characteristic radius $R$ and total mass of stars $M_{*}$.
This corresponds to the gravitational potential
\begin{equation}
    \Phi(r) = -\frac{G M_*}{R \sqrt{1+  \frac{r^2}{R^2}}}
\end{equation}
and mass enclosed in a sphere of radius $r$:
\begin{equation}\label{eq:Mofr}
    M(r) = \frac{M_{*} r^3}{R^3}\left( 1 + \frac{r^2}{R^2}\right)^{-3/2}.
\end{equation}
A detailed description of how positions and velocities of particles are determined is give in Appendix \ref{app:PlummerICs}. 
Particles chosen by this scheme are near a stable configuration in an external potential but are not entirely stable. We therefore simulate them in an external NFW potential with the same parameters as the dark matter halo until the distribution has reached an equilibrium. For simulations of Plummer spheres in plane-wave boxes it is not necessary to also include the contribution of a halo potential.


\textbf{Test particles in circular orbits.}
In these simulations, massless point test particle is assigned a random radius, $|r_i| \in [0, R_\mathrm{max}]$, and then 
given velocity consistent with a circular orbit around the center of a halo but random orientation in three dimensions. 
The algorithm to assign position and velocity of particles is as follows.
\begin{enumerate}
    \item We assign a particle radius $R_i$.
    \item Pick three additional random variables from a normal distribution with unit variance, i.e. $Y_1, Y_2, Y_3 \sim \mathrm{Normal}$.
    \item The particle position is then $\vec r_i = R_i \frac{Y_1 \hat x + Y_2 \hat y + Y_3 \hat z}{\sqrt{Y_1^2 + Y_2^2 + Y_3^2}}$.
    \item Next we write the unit position vector, $\hat r_i = r_i / R_i$, and an arbitrary orthogonal unit vector, $\hat o_i$ (for example $\hat o = \hat r_y \hat x - \hat r_x \hat y$).
    \item We then find two orthogonal unit vectors in the tangent plane, e.g. $\hat t_i = \hat r_i \times \hat o_i$ and $\hat b_i = \hat t_i \times \hat r_i$.
    \item Pick a random unit vector in this tangent plane using a uniform random angle $\theta_i \in [0,2\pi)$, i.e. $\hat v_i = \hat t_i \sin(\theta_i) + \hat b_i \cos (\theta_i)$.
    \item Normalize the velocity by the circular velocity at that radius using the total enclosed mass of the dark-matter halo and stars. 
    $M(<R_i)$, i.e. $\vec v_i = \sqrt{\frac{G \, M(<R_i)}{R_i}} \hat v_i$.
\end{enumerate}

\textbf{Free particles with a Maxwell-Boltzmann velocity distribution}
In the plane-wave case we consider free particles whose position are initially uniformly randomly distributed in the box and have velocities following Maxwell-Boltzmann distribution 
\begin{equation}
f(\mathbf{v})\,d^{3}\mathbf{v}=\left({\frac{1}{2\pi \sigma_*^2}}\right)^{3/2}\exp\left(-{\frac{ v^{2}}{2\sigma_*^2}}\right)\,d^{3}\mathbf{v}.
\end{equation}
Or as a probability distribution of the speeds
\begin{equation}
f(v)=\left({\frac{1}{2\pi \sigma_*^2}}\right)^{3/2} 4\pi v^2 \exp\left(-{\frac{ v^{2}}{2\sigma_*^2}}\right)
\end{equation}

To obtain the velocity vector for each star we therefore draw three normally-distributed numbers $u_x,u_y,u_z$ and multiply them by $\sigma_*$ to obtain a vector whose elements follow Maxwell-Boltzmann distribution: $\mathbf{v} = (v_x, v_y, v_z) = \sigma_*(u_x, u_y, u_z)$ with $v_x,v_y,v_z$.

\subsection{Solver}

We integrate the equations of motion using the standard symplectic leap frog solver on a fixed resolution grid. Schematically the solver works as follows
\begin{enumerate}
    \item Half step position update,
    \item Full step momentum update,
    \item Half step position update.
\end{enumerate}
A half-step position update for the particles and the drift for the field is the following for each field and particle, respectively,
\begin{align}
    &\psi_i(t + \Delta t) = \mathcal{F}^{-1}\left[ e^{-i\hbar k^2 \Delta t / 4m_i} 
    \mathcal{F}\left[ \psi_i(t) \right] \right] \, , \\
    &\vec r_j(t + \Delta t) = \vec r_j(t) + \Delta t \, \vec v_j(t) /2  \, .
\end{align}
Where $\mathcal{F}$ is a Fourier transform. A momentum update for the particles and the kick for the field is given as follows

\begin{align}
    &\psi_i(t + \Delta t) = e^{-i m_i V_1(x_g) \Delta t / \hbar} \psi_i(t) \, , \\
    &\vec v_j(t + \Delta t) = \vec v_j(t) + \Delta t \, \vec a(r_j)
\end{align}
\begin{widetext}
Where $x_g$ is the fixed resolution grid on which our fields, $\psi_i$, are defined.  The various potentials and acceleration above are given by
\begin{align}
    &\nabla^2 V_{*}^{\mathrm{CIC}}(x_g) = 4 \pi G \, \rho^{\mathrm{CIC}}_* \, , \\
    &\nabla^2 V_{\mathrm{ULDM}}(x_g) = 4 \pi G \sum_i |\psi_i(x_g)|^2  \, , \\
    &V_1(x_g) = V_{\mathrm{NFW}}(x_g) + V_{*}^{\mathrm{CIC}}(x_g) + V_{\mathrm{ULDM}}(x_g) \, , \\ 
    &\vec a(r_j) = -(\nabla_5 V_{\mathrm{ULDM}})|_{r_j} - r_j  \frac{G \, M_{\mathrm{NFW}}^{\mathrm{end}}(r_j)}{|r_j|^3} - \sum_k (r_k - r_j) \frac{G \, m_k }{|r_k - r_j + \epsilon|^3}
\end{align}
Where the CIC superscript indicates that a density was calculated using a cloud in cell deposit scheme. $m_k$ indicated the mass of the $k$th particle. $\nabla_5$ correspond to a 5-pt first derivative stencil, $|_{r_j}$ means the value is linearly interpreted at the position $r_j$. $\epsilon$ represents a gravitational softening term and is set to be small compared to the grid cell size. The NFW potential, $V_{\mathrm{NFW}}$, and enclosed mass, $M_{\mathrm{NFW}}^{\mathrm{end}}$, are given by analytic expressions below.
\begin{align}
    V_{\mathrm{NFW}}(r) &= -4 \pi G \rho_0 R_s^3 \, \log(1 + r/R_s) / r \, \label{eq:NFWpot}\\
    M_{\mathrm{NFW}}^{\mathrm{enc}} &= 4 \pi \rho_0 R_s^3 \left( \log((R_s +r)/R_s) - r / (R_s + r)\right) \, .
\end{align}
\end{widetext}
\section{Results} \label{sec:results}

\subsection{No equilibrium solution}
\begin{figure*}[!ht]
	\includegraphics[width = .67\textwidth]{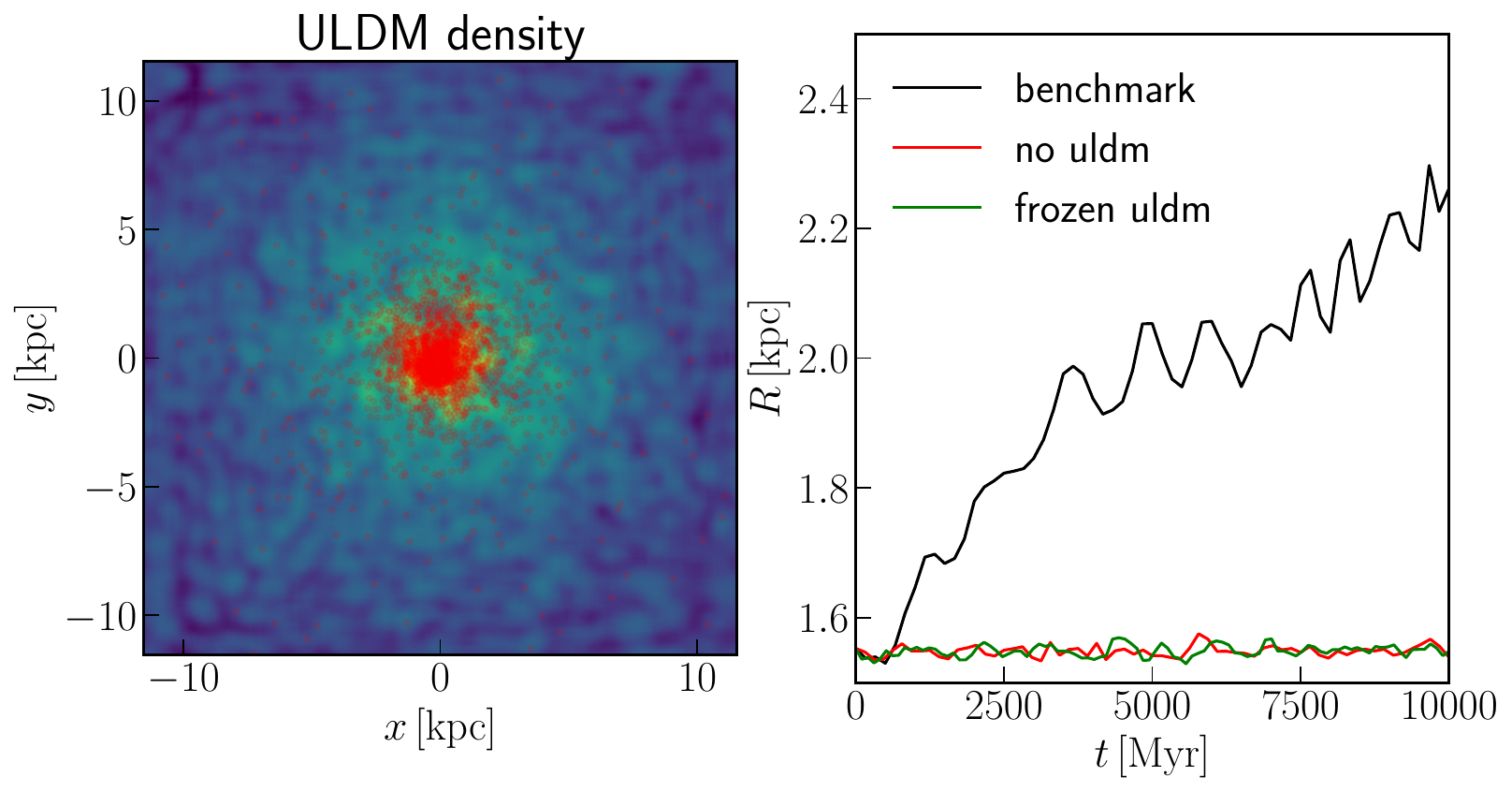}
	\caption{ \textbf{Left:} the log density of an ultralight dark matter halo. The red dots overlayed on the density show the projected position of stellar particles. \textbf{Right.} Change in half-light radius of stars embedded in evolving ultralight dark matter is plotted (black). This is compared with the same distribution of stars evolving in the analogous NFW profile (red) and in a ``frozen" ultralight dark matter halo (green). The latter two simulations do not experience any heating or change in the half-light radius.  }
	\label{fig:heating_general}
\end{figure*}

The heating of stellar particles embedded in an ultralight dark matter halo is a fairly robust result. The gravitational scattering of stellar particles with density fluctuations results in a transfer of energy to the stars resulting in larger orbits and half-light radius. Our simulations of stars distributed in evolving ultralight dark matter halos have shown an increase in the half-light radius over time, see Figure \ref{fig:heating_specific}. This effect is consistent with previous works \cite{Dalal2022, DuttaChowdhury2023, Marsh:2018zyw, Hui_2017} (in Appendix \ref{app:segueii} we reproduce the heating found in \cite{Dalal2022}). 

In principle one could argue that the stars, which are in equilibrium in an NFW halo, might relax due to transient effects, once placed in an ULDM halo, to a new equilibrium in a way that mimics heating. However, we show the heating is caused by the interaction of the stars with the evolving ultralight dark matter halo as opposed to a simple relaxation to a new equilibrium distribution with a larger half-light radius. To demonstrate this we compare the results of three simulations, plotted in Figure \ref{fig:heating_general},
\begin{enumerate}
    \item \textbf{Benchmark:} A simulation with a distribution of stars set in a dynamically evolving ultralight dark matter halo. The results for this simulation are shown in Figure \ref{fig:heating_specific}. The stars half-light radius increases, consistent with heating from interactions with the ultralight dark matter halo. The mass of the particle is $m_{22} = 5$ and the ULDM fraction is $f_\mathrm{ULDM} = 0.5$.
    \item  \textbf{No ultralight dark matter:} The stellar distribution in this simulation is the same as the benchmark. However, the ultralight dark matter fraction is $0$. Stars evolve in an static, spherically symmetric NFW profile. 
    The stellar distribution is in equilibrium and does not heat over time. This demonstrates that our initial stellar configuration is stable in the absence of ULDM fluctuations.
    \item \textbf{Frozen ultralight dark matter:} In this simulation the stellar distribution and ultralight dark matter halo are the same as the benchmark. However, the ultralight dark matter substructure is static and does not evolve. The ultralight dark matter granules and soliton are still present but stationary. We can see in this case that the stellar distribution does not heat and remains in its initial equilibrium.
\end{enumerate}
These simulations clearly show that (a) the initial stellar distribution is stable on the timescales considered in this work, (b) the heating comes from the dynamical nature of the granules, not simply from stars tracing an inhomogeneous density field, and (c) in the presence of dynamical fluctuating ULDM density stars do no relax into an equilibrium configuration on the relevant time scales.


\subsection{Heating using quasi particle approximation} \label{subsec:quasiParticleApprox}
\begin{figure*}[!ht]
	\includegraphics[width = .67\textwidth]{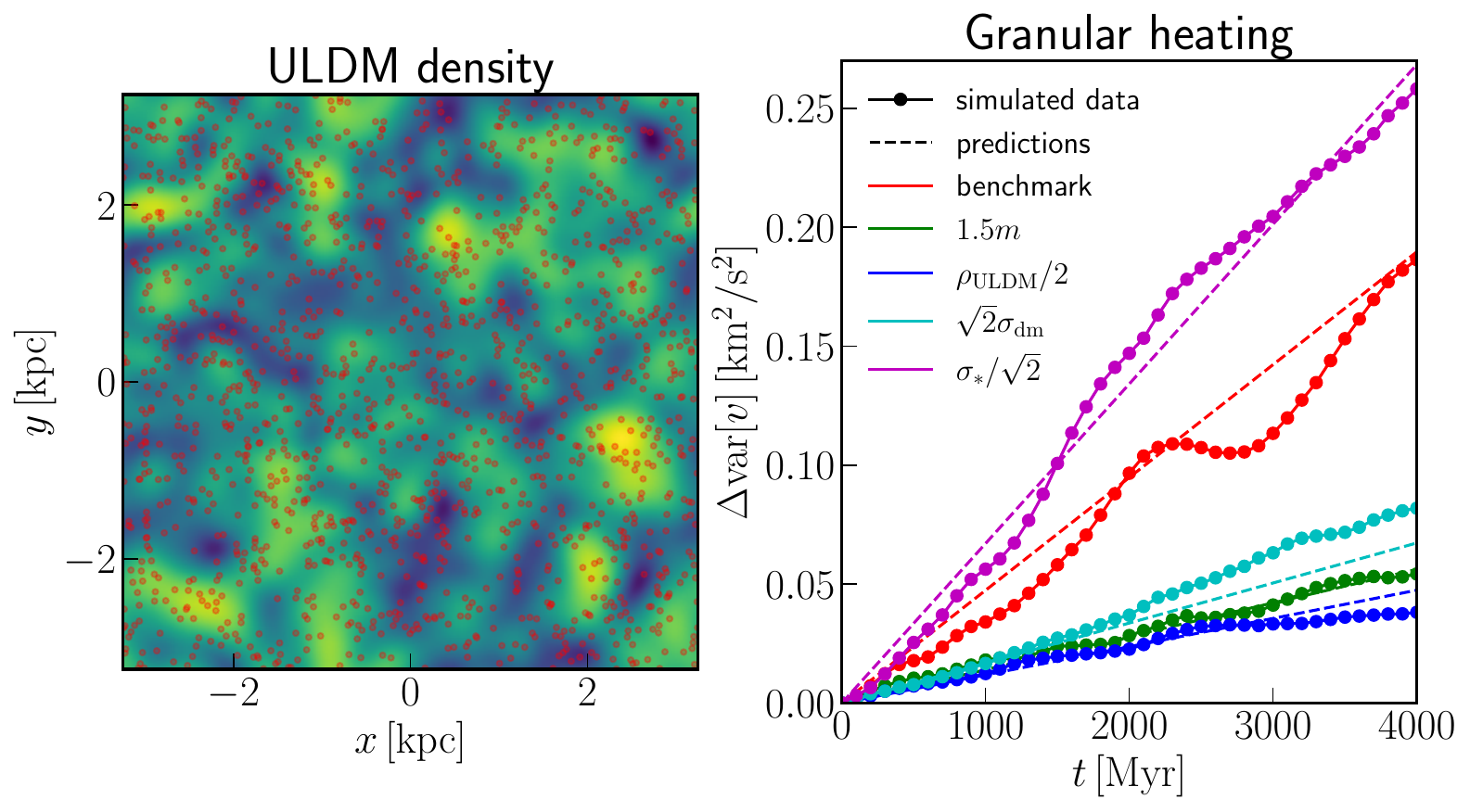}
	\caption{ \textbf{Left:} the log density of an ultralight dark matter plane-wave box. The red dots plotted over the density show the projected position of stellar particles. \textbf{Right.} The results of a set of plane-wave box simulations. The increase in the variance of the particle velocity distributions is plotted as a function of time. Solid lines correspond to simulated data and dashed lines to the corresponding variance change prediction using equation \eqref{eqn:deltaVar}. Different colors correspond to different simulation parameters.     
    The predictions reliably predict the change in the variance for the simulations shown. The particles are initialized with velocity dispersion, $\sigma_*$, uniformly positioned in the box. Here $m$ refers to the ULDM field mass, $\rho_\mathrm{ULDM}$ refers to the ULDM density, and $\sigma_\mathrm{DM}$ the dark matter velocity dispersion. The line labels given in the figure
    describe how the relevant simulation parameters are changed with respect to the benchmark values.  
    For instance $\sigma_*/\sqrt{2}$ means $\sigma_*$ is decreased by a factor of $\sqrt{2}$ from its benchmark value.
    }  
	\label{fig:granuleHeatingFree}
\end{figure*}

\subsubsection{Plane-wave box simulations}

We study the scaling of this effect using the planewave box simulations described in Section \ref{sec:simICs}. A superposition of planewaves has been found to describe the density fluctuations in halos for regions small compared to the dynamical length scales of the halo \cite{Dalal_2021}, and this approximation has been applied to similar works\cite{Eberhardt2025, Eberhardt2024}. Here, particles are initialized with random positions in the simulation box and their velocities are drawn from Boltzmann distribution with parameter $\sigma_*$.
We find that the quasiparticle approximation predicts the change in the velocity variances well, i.e., equation \eqref{eqn:deltaVar}.
 Here we assume the relative velocity is given by the stellar velocity dispersion $v_r \sim \sigma_*$. We test a number of systems varying the value of $\rho_\mathrm{ULDM}$, $m$, $\sigma_*$, and $\sigma_\mathrm{DM}$ and track the change in the velocity variance for an ensemble of free particles in the box. 
The results are plotted in Figure \ref{fig:granuleHeatingFree} comparing simulated data and prediction. We can see that equation \eqref{eqn:deltaVar} reasonably approximates the change in the velocity variance. 

\begin{figure}[!ht]
	\includegraphics[width = .44\textwidth]{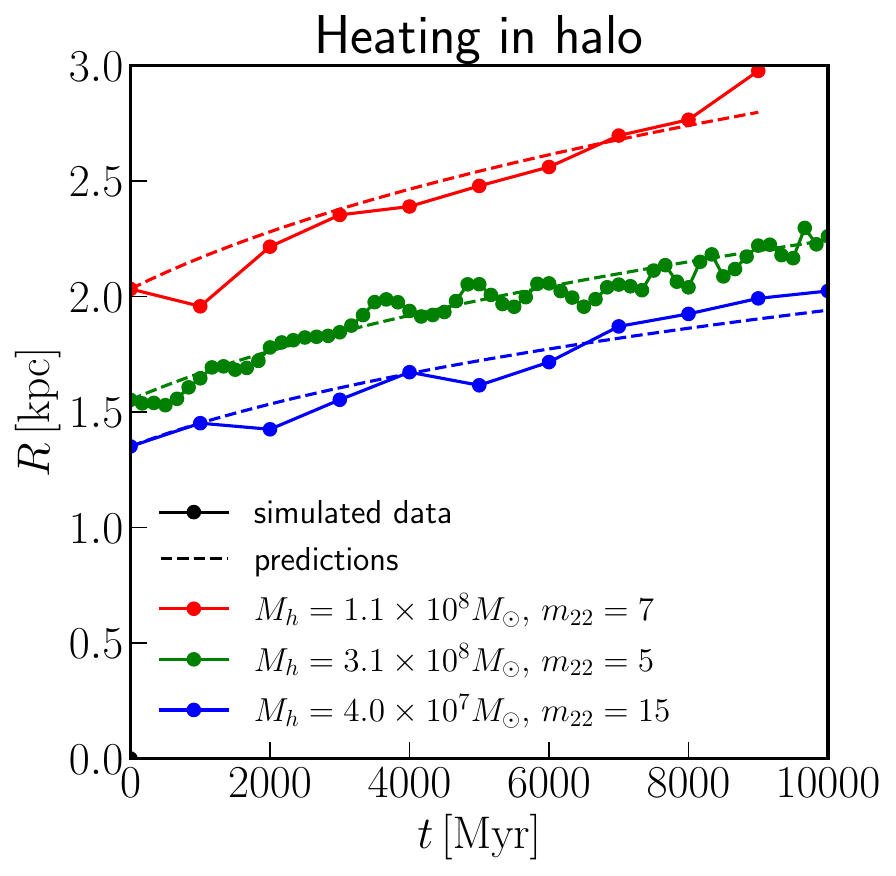}
	\caption{ Change in half-light radius for stellar distributions in a few different ultralight dark matter halos with a variety of halo and particle masses. Importantly, the half-light radius of the stars in these simulations is large compared to the core radius, which for these halos is $\lesssim 0.5 \, \mathrm{kpc}$. The solid lines show the simulated data for each halo and the corresponding dotted line of the same color shows the prediction using the quasi-particle approximation applied to these halos, i.e., equation \eqref{eqn:Rhalf_predict}. We can see that the prediction is accurate over the course of the simulations in each halo tested. }
	\label{fig:haloHeating}
\end{figure}

\subsubsection{Halo simulations, large $R$} \label{subsec:haloHeating_skirt}
In this section we simulate particles in a virialized distribution within a halo. These initial conditions are described in Section \ref{sec:simICs}. Here, we will discuss systems where the half-light radius is large compared to the core and so the system should be well described by our analysis in Section \ref{sec:farFromCore}. 

In the halo, the particles are on orbits, this means that the density and size of ultralight dark matter granules they encounter may not be well described by a plane wave box with a single velocity variance parameter. In Figure \ref{fig:haloHeating} we compare the growth of the half-light radius of the simulated data to the prediction using equation \eqref{eqn:dr_predict_largeR}. The dashed lines indicates the result of the integration of equation \eqref{eqn:dr_predict_largeR} and the solid lines the actual half-light radius of the stellar distribution. The agreement is quite good. This implies that the quasi-particle approximation is predictive for stellar distributions in the outer halo.


\subsection{Soliton heating} \label{sec:solitonHeating}

\begin{figure*}[!ht]
	\includegraphics[width = .99\textwidth]{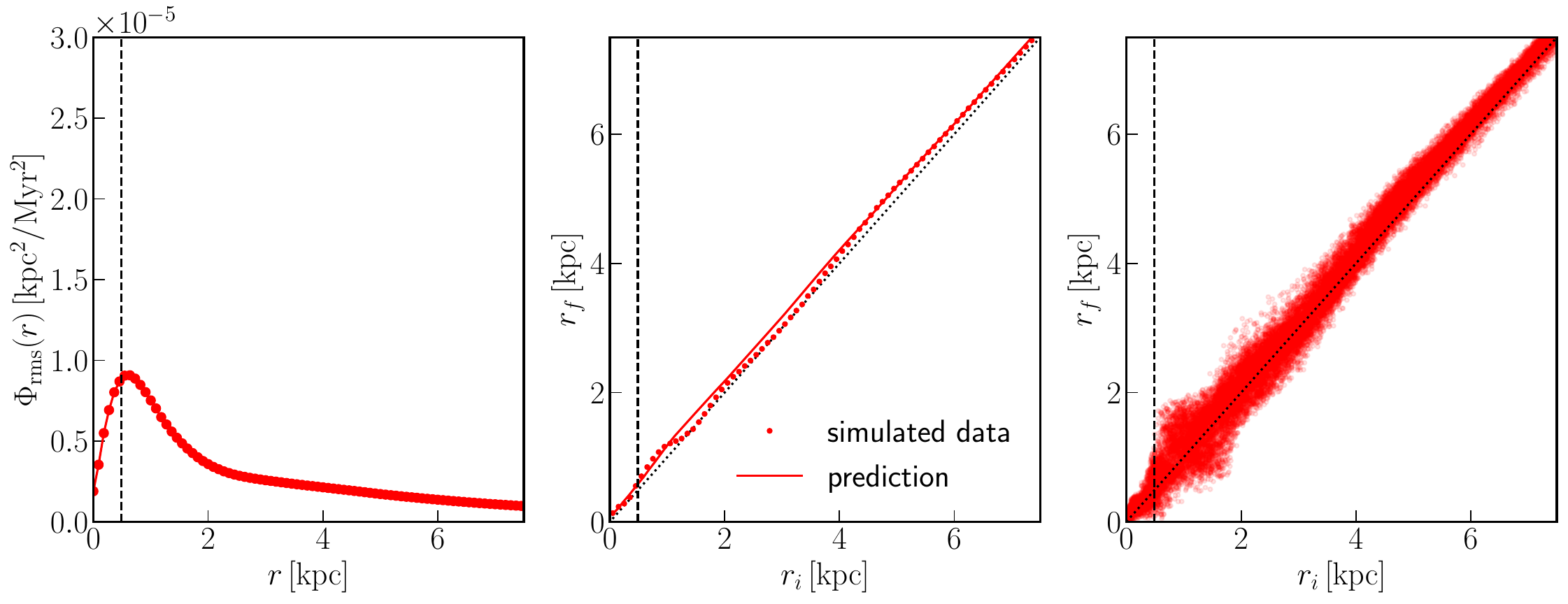}
	\caption{ \textbf{Left:} The root-mean-square of the potential fluctuations (see equation \eqref{eqn:phi_rms}), averaged over time and angle, as a function of radius as measured in a simulation. The potential fluctuations are relatively large within a few core radii and then fall off outside this region. This is due to the random walk and oscillation of the central soliton. The dashed black line in the left plot shows the core radius. \textbf{Center:} A plot of the average final radii for particles in initially circular orbits as a function of the initial orbital radius. The solid line indicates the expected average value given the rms potential fluctuations (show in the left plot). We can see that this is fairly predictive especially inside a few soliton radii. The dashed black line indicates $r_i = r_f$ or not average change. We can see that most points are above the dashed black line consistent with our expectation that the stars should experience heating. \textbf{Right:} The initial final radii plotted for each particle. In addition to sending the particles to larger orbits on average the potential fluctuations also induce a spread in the orbital parameters. The large disruption in the inner region corroborating the idea that the motion of the soliton creates a large potential disruption. The data here is for a field of mass $5 m_\mathrm{22}$, the snapshot above corresponds to a simulation time of $1 \, \mathrm{Gyr}$.  }
	\label{fig:dr_phi_rms}
\end{figure*}

\begin{figure*}[!ht]
	\includegraphics[width = .67\textwidth]{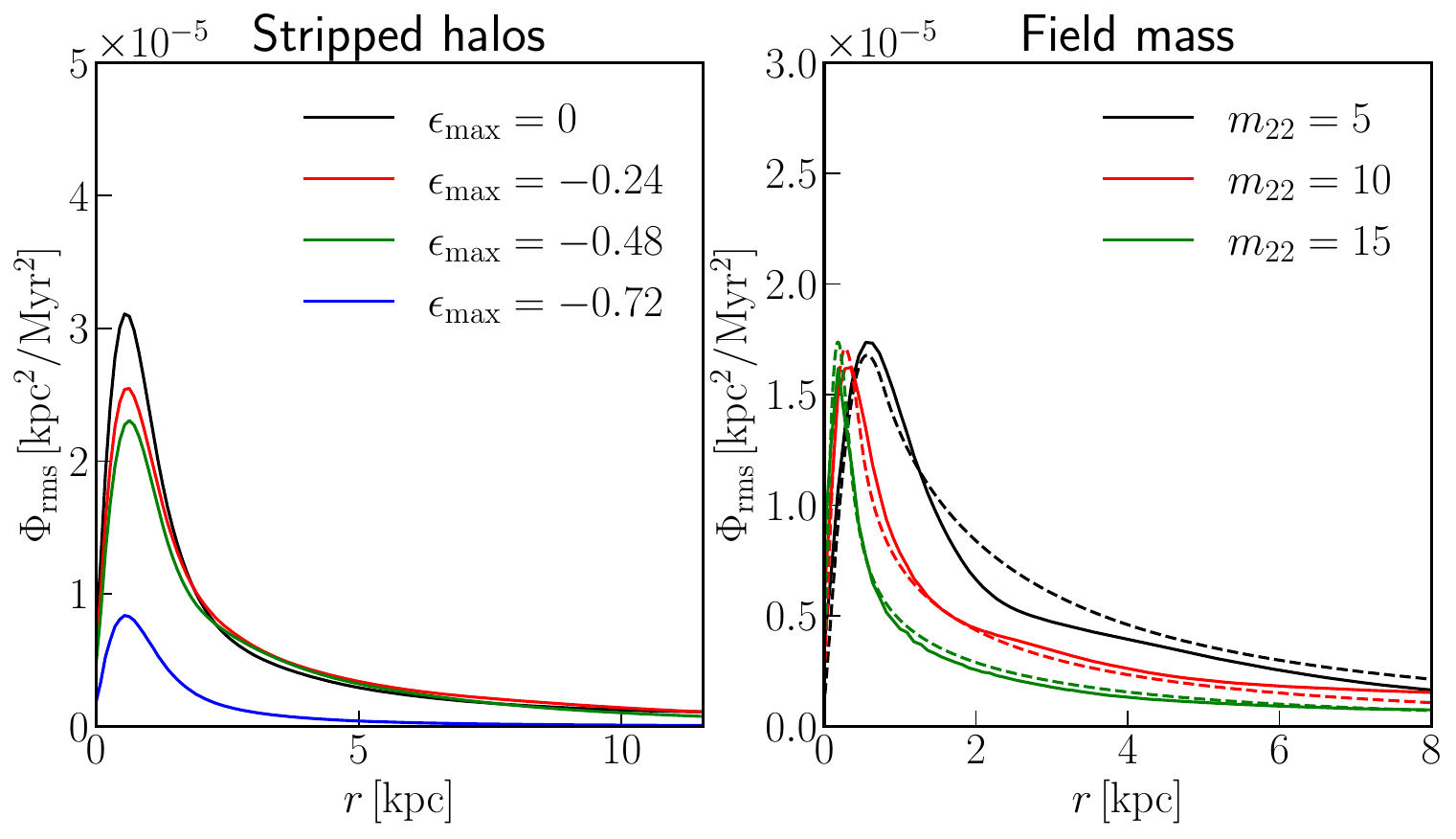}
	\caption{ The average root-mean-square potential fluctuations over the spherically averaged radial potential. \textbf{Left:} the potential fluctuations for halos with differing degrees of stripping. We can see that stripping does little to affect the potential fluctuations except in the most stripped case where the tidal radius is of order the de Broglie wavelength. \textbf{Right:} potential fluctuations for halos with different field masses (solid lines) and corresponding predictions (dashed lines) using equation \eqref{eqn:phi_rms}. We can see that the prediction accurately describes the potential fluctuations in the central soliton region.  }
	\label{fig:phi_fluctuations}
\end{figure*}

\begin{figure}[!ht]
	\includegraphics[width = .44\textwidth]{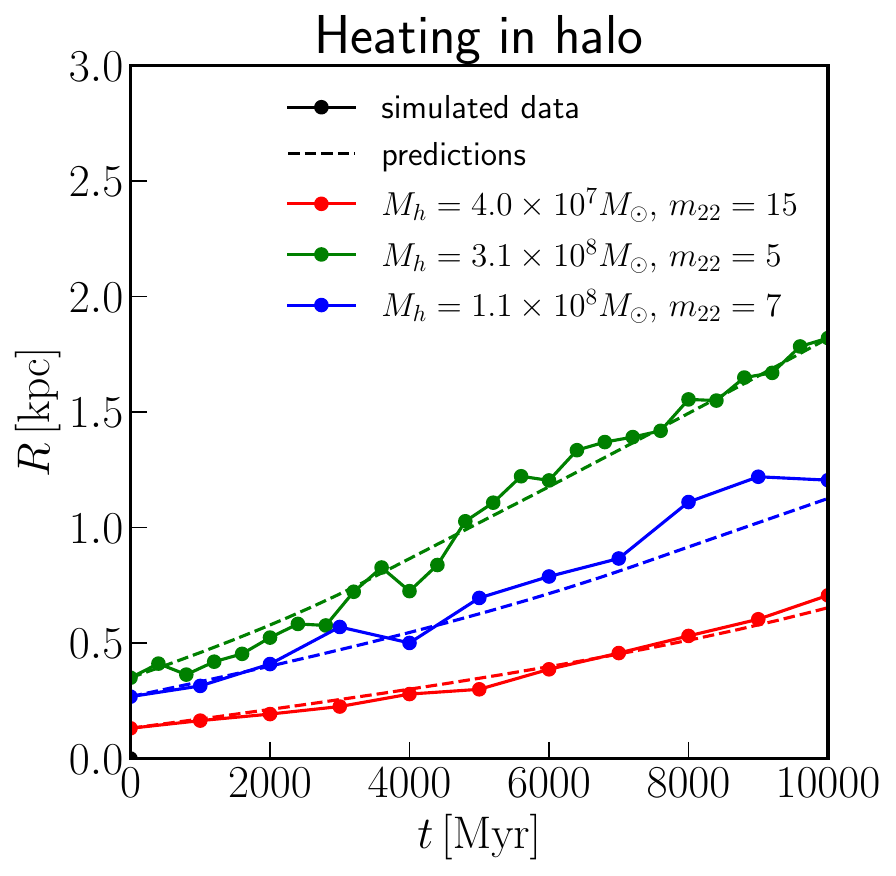}
	\caption{ Change in half-light radius for stellar distribution in a few different dark matter halos. Importantly, in these simulations, the initial half-light radius is within the core which for these halos is $\lesssim 0.5 \, \mathrm{kpc}$. The solid lines show simulated data for each halo and the corresponding dotted line of the same color shows the prediction using the expected potential fluctuations. We see the predictions are reasonably accurate, particularly at early times when the stars are within the core.  }
	\label{fig:solitonHeatingHalo}
\end{figure}
 
In this Section we simulate the heating due to the soliton oscillation and random walk. We first test the heating as a function of the orbital radius. To do this we start an ensemble of test particles in randomly-oriented circular orbits centered on the halo. The heating due to the granules and soliton will cause the average orbital radius of the ensemble to increase overtime. The results of this test are plotted in Figure \ref{fig:dr_phi_rms}. The left panel shows the root-mean-square potential fluctuations (equation \eqref{eqn:phi_rms_define}).

These measured potential fluctuations (left panel of Figure \ref{fig:dr_phi_rms}), $\Phi_\mathrm{rms}$, predict a change in energy (equation \eqref{eqn:delta_E_soliton}) which corresponds to a change in the average orbital radius of particles (equation \eqref{eqn:f_define}). We plot the final average orbital radius as a function of initial orbital radius for the simulated data compared with this predicting in the center panel of Figure \ref{fig:dr_phi_rms}. We can see that the change in radius is reasonably predicted. Therefore, we conclude that the potential fluctuations can be used to predict the heating of the stellar distribution. But we will test this further below.

Therefore, in order to predict the heating in distributions near the soliton we need to be able to reliably predict $\Phi_\mathrm{rms}(r)$. We plot measure $\Phi_\mathrm{rms}(r)$ in a number of different simulations varying the field mass. We compared the measured $\Phi_\mathrm{rms}(r)$ (solid lines) with equation \eqref{eqn:phi_rms} (dashed lines)  in the right panel of Figure \ref{fig:phi_fluctuations}. We obtain a reasonable fit, particularly within a few core radii. We point out, that equation \eqref{eqn:phi_rms} is a phenomenological fit to the data. 

Now that we have some confidence that 1) the potential fluctuations can predict the change in energy, and 2) we can predict the potential fluctuations near the soliton, we move on to simulations of virialized stellar distributions inside actual halos. We test the full prediction for the change of the half-light radius near the solitonic core, i.e. the equation \eqref{eqn:Rhalf_soliton}. We test this prediction by simulating stellar distributions in ultralight dark matter halos with initial half-light radii within the soliton. The results are plotted in Figure \ref{fig:solitonHeatingHalo} for a few different halos. Predictions performed by integrating equation \eqref{eqn:delta_E_soliton} are plotted as dashed lines and the corresponding data as solid. We can see that this equation is quite predictive, particularly at early times when the stars are well within the region affected by the soliton.

\subsection{Stellar self-potential}

\begin{figure}[!ht]
	\includegraphics[width = .44\textwidth]{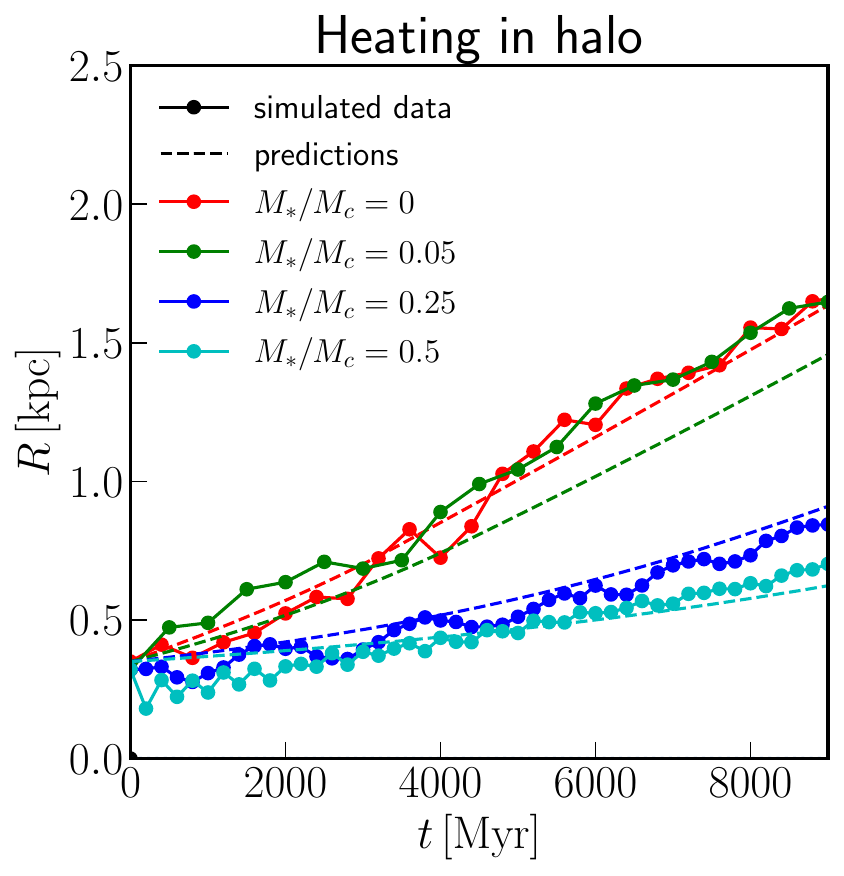}
	\caption{ The change in the half-light radius for particles in a dark matter halo for a few different stellar masses. In these simulations the initial half-light radius is slightly smaller than the core radius, $R \lesssim r_c$. t In the legend we show the total stellar mass, $M_*$, as a fraction of the core mass, using the core-mass relation found in \cite{Schive2014}. We can see that for small ratio the heating is unaffected by the stellar self-potential. However, when the stellar mass is similar to the core mass, we can see that the change in the half-light radius is substantial. }
	\label{fig:selfPotential}
\end{figure}

In this section we will discuss simulations in which we vary the stellar particle masses, keeping the ultralight dark matter halo fixed. In other sections the particle masses has been chosen such that the stellar density is small compared to the dark matter density. Here we study the effect of relaxing this assumption. Results are plotted in Figure \ref{fig:selfPotential}. Notice that we can still predict the correct change in the half-light radius by including the stellar self potential, $E_* \sim G M / R$, in equation \eqref{eqn:totalEnergy} when calculating the energy. Importantly, we have only included the stellar self-potential in this updated prediction. We have not modeled the back-reaction of the stellar potential on the ULDM halo which may add additional corrections to the heating rate, although from our results they appear not to be significant in the systems we studied.

One would expect more massive stars to be more tightly bound and therefore experience a smaller change in their half-light radius given the same amount of energy transferred from the ultralight dark matter halo. We can see if we increase the mass of the stars sufficiently that the half-light radius indeed grows at a much slower rate. When the stellar potential needs to be comparable to the dark-matter potential the effect is significant. When we increase the stellar particle mass but keep the total mass of the stars small compared to the halo core they are in, the half-light radius change was consistent with the massless limit. 

\subsection{Tidal stripping}
\begin{figure*}[!ht]
	\includegraphics[width = .97\textwidth]{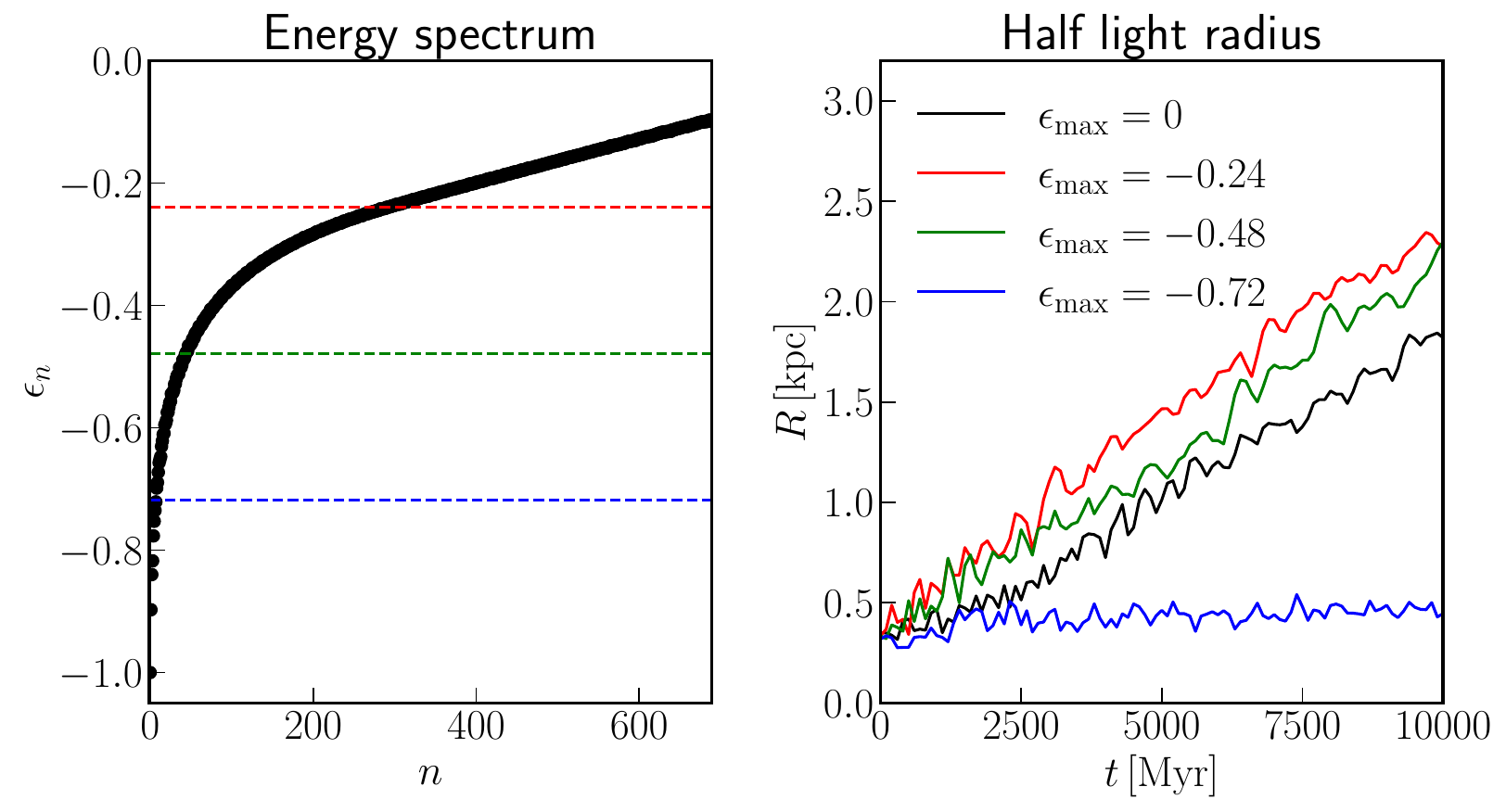}
	\caption{ \textbf{Left:} the energy spectrum of the dark matter halo. $\epsilon \equiv E / |E_0|$ is the energy as a fraction of the ground state energy, $n$ is the eigen number of each mode. The dashed line show the different energy thresholds used to construct stripped halos. \textbf{Right:} The heating resulting from each stripped halo compared with the benchmark simulation ($\epsilon_\mathrm{max} = 0$, black) with all eigenmodes included. We quantify the heating using the fractional change in the half-light radius of the particles. We can see that the heating of the stars in the halo is largely unchanged except for the case with the most substantial stripping (blue). }
	\label{fig:spectrum_and_heating}
\end{figure*}

In this section we will discuss the effect of tidal stripping on the heating rate. We create tidally stripped halos by truncating the energy eigenmodes used in the halo construction as discussed in Section \ref{subsec:ULDM_ICs}. Because of the truncation, the newly constructed halos are not in exact equilibrium. We simulate the newly constructed halos for $1 \, \mathrm{Gyr}$ until they settle into an equilibrium state. The halos resulting from this procedure are plotted in Figure \ref{fig:strippedHalos}. We then simulate a stellar distribution in the new halo with the field evolution frozen in time until the stellar distribution settles into equilibrium as well, which also takes about $1 \, \mathrm{Gyr}$. After that, allow the ULDM field to begin evolving.

We plot the change in the half-light radius over time in halos with differing degrees of stripping in Figure \ref{fig:spectrum_and_heating}. In the left panel of this figure we show the energy eigenspectrum of the original halo. The colored dashed lines correspond to the truncation energy used to generate each halo. In the right panel we show the change in the half-light radius for stars placed in each of the halos. The black line corresponds to the benchmark simulation in which the energy eigenmodes are not further truncated. The red, green, and blue lines correspond to varying degrees of stripping from the least to the most stripped, respectively. We can see that the heating of the stars in the two least stripped halos (red and green) is very close to the heating in the benchmark (black) simulation. However, the heating of stars in the most stripped halo (blue) is significantly reduced. 

Significant stripping can substantially reduce the heating. We can see in Figure \ref{fig:strippedHalos} that the most stripped halo has a tidal radius of order the de Broglie wavelength. In this case we see substantially reduced heating, for situations where the tidal radius is large compared to the de Broglie wavelength the heating is unaffected. This analysis is corroborated by looking at how the potential fluctuations change as a function of the stripping parameter. This is done in the left panel of Figure \ref{fig:phi_fluctuations}. We can see the potential fluctuations barely change until the halo is extremely stripped consistent with the similar observed heating rates shown in Figure \ref{fig:spectrum_and_heating}. 

It is important to note however, that the effect of stripping on heating as we have modeled it here is not monotonic. We can see if Figure \ref{fig:spectrum_and_heating} that the slightly stripped halos (red and green) actually under greater heating than the benchmark simulation. This is due to the fact that we are not fixing the inner density of the halo and after our stripping procedure the halo will relax into a new equilibrium in which the inner halo density may differ. However, by the time the halo is extremely stripped (blue), the inner density is less relevant since the central soliton barely oscillated or randomly walks at all.

\subsection{Dark matter fraction in outer halo}
\begin{figure}[!ht]
	\includegraphics[width = .44\textwidth]{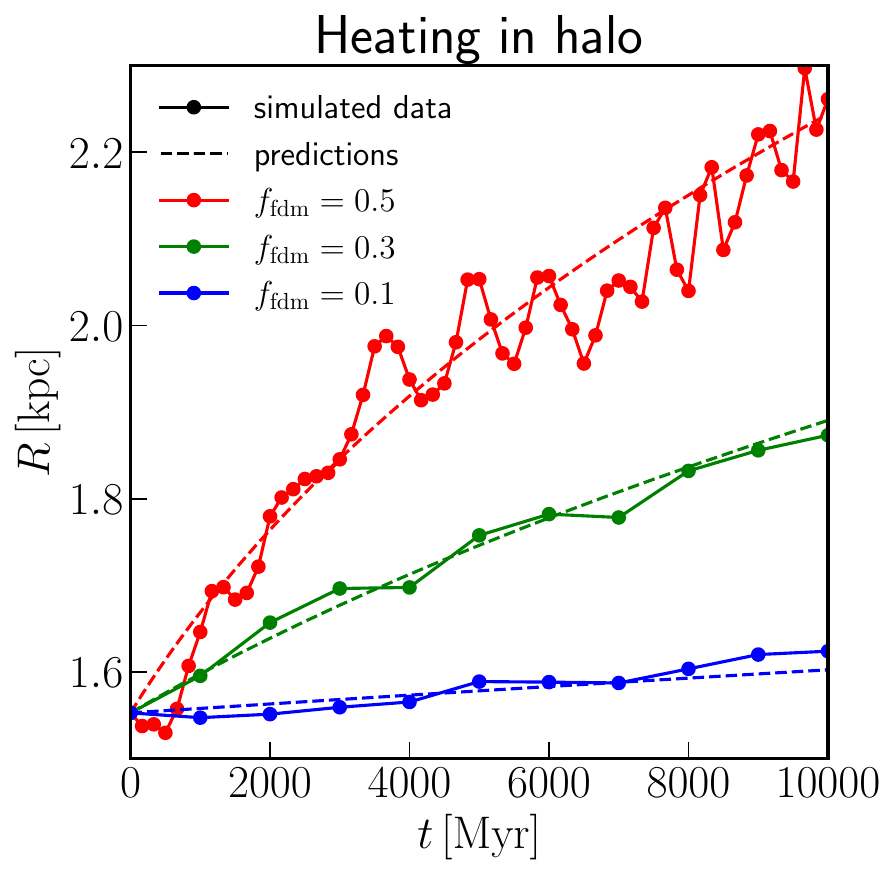}
	\caption{ Halo simulations showing the heating of stellar distributions for a few different ultralight dark matter fraction. The results are consistent with the quadratic scaling presented in equation \eqref{eqn:dr_predict_largeR}.}
	\label{fig:varyingFraction}
\end{figure}

It has been argued that the dynamical heating of stars in the outer halo should be $\propto f_\mathrm{ULDM}^2$, as predicted by the quasi-particle approximation \cite{Dalal2022, Gosenca2023, teodori2025, Hui:2021tkt}. It has also been argued that the scaling should be $\propto f_\mathrm{ULDM}$ \cite{Marsh:2018zyw}. Here we simulate a few halo varying the total fraction of the ultralight dark matter. We plot the change in the half-light radius against predictions assuming that the correct scaling is $\propto f_\mathrm{ULDM}^2$ in Figure \ref{fig:varyingFraction}. We can see that the $\propto f_\mathrm{ULDM}^2$ scaling is very predictive implying that this is likely the correct scaling with ultralight dark matter density.

\subsection{Heating with and without the soliton} \label{sec:discussion}

\begin{figure}[!ht]
	\includegraphics[width = .47\textwidth]{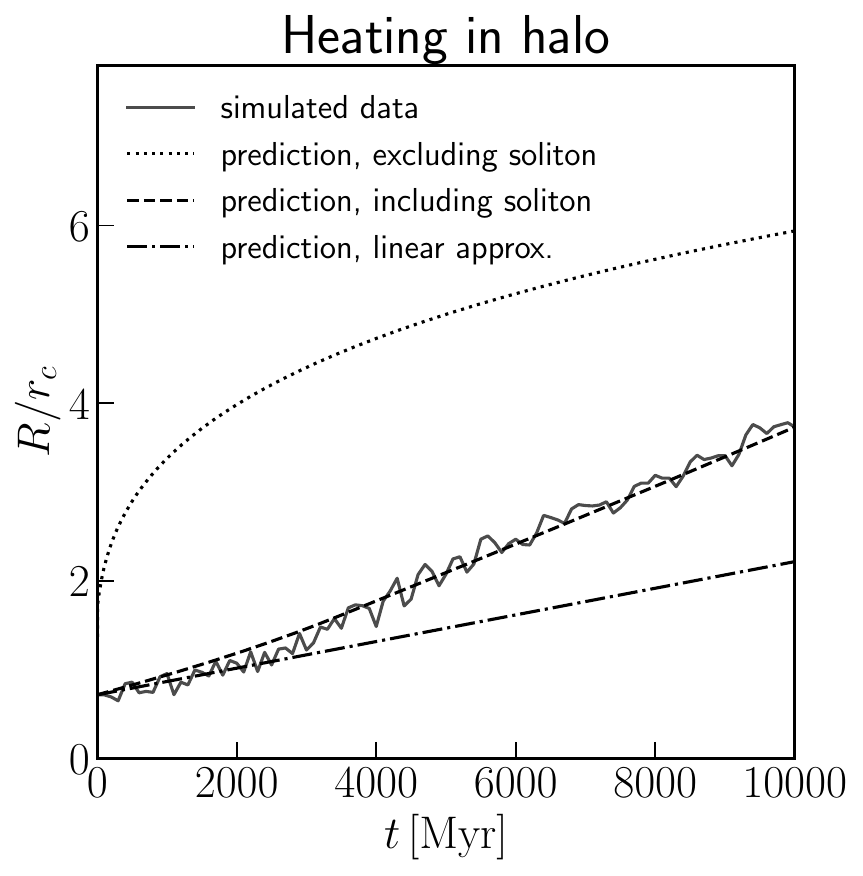}
	\caption{ The change in the half-light radius for stellar particles simulated in a halo. The initial half-light radius is within the core radius. The data are compared with three different analytic predictions. The first, shown as a dotted line, is the quasi-particle approximation discussed in Section \ref{subsec:quasiParticleApprox}. We can see that this approximation is not accurate, especially at early times when the approximations that go into it are not well motivated. The second approximation, shown as a dashed line, is the heating prediction including the impact of the central core; this was discussed in Section \ref{sec:solitonHeating}. This prediction matches the data much more closely. Finally, we plot the simple linear approximation (equation \eqref{eqn:linearApprox}), shown as a dash-dot line. We can see that this approximation works well only at early times, consistent with expectation.
    }
	\label{fig:comparePredictions}
\end{figure}

\begin{figure}[!ht]
	\includegraphics[width = .47\textwidth]{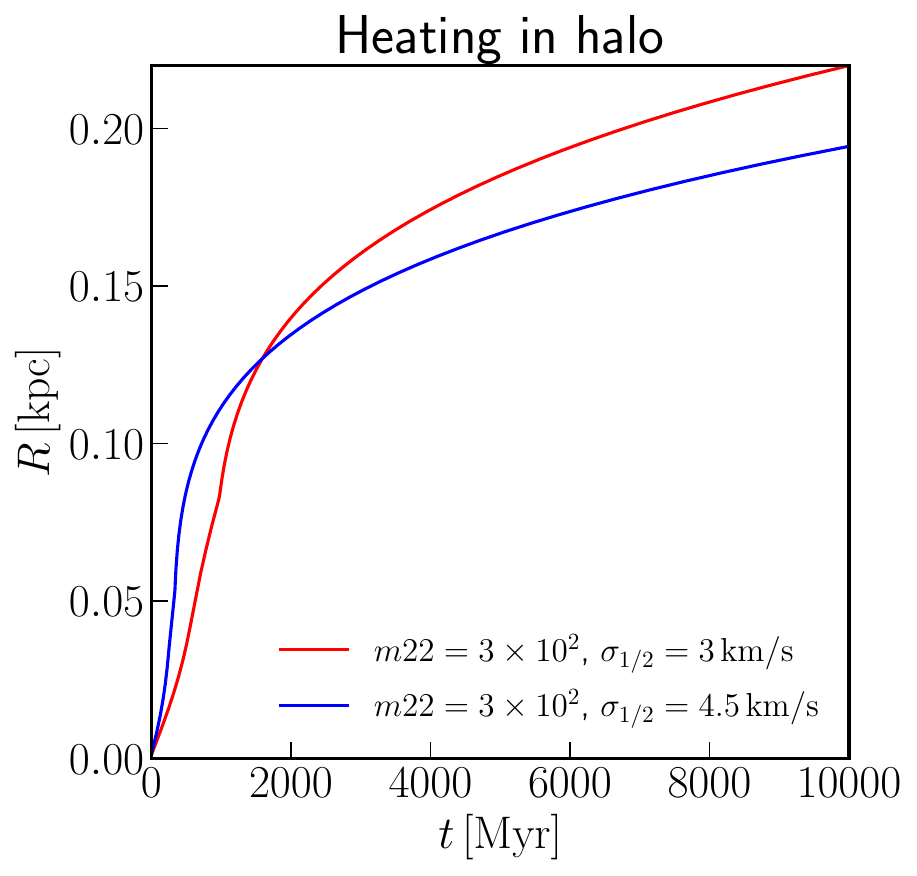}
	\caption{ Predictions of the change in the half-light radius for two halos with different velocity dispersions. We can see that the results are affected in the regime where the soliton heating is important and in the outer halo.  }
	\label{fig:compareDispersion}
\end{figure}

In Figure \ref{fig:comparePredictions} we plot the change in the half-light radius for a simulated halo. We plot the simulated data compared with three different predictions. The first, plotted as a dash-dot line, is the linear prediction for the growth of the half-light radius when the stars are within the central core, i.e., equation \eqref{eqn:linearApprox}. The second, plotted as a dotted line, is the prediction assuming that the soliton radius is 0, or in other words, that the quasi-particle approximation is valid at all times, i.e., equation \eqref{eqn:dE_far}. The third, plotted as a dashed line, is the prediction using our expression for soliton heating, equation \eqref{eqn:dE_near}, when the half-light radius is of order the soliton radius, and switching  to the quasi particle approximation, only when the half-light radius is large (we use about a factor of $R\sim 4r_c$) compared to the core radius. 

As expected we can see that the linear approximation is only good at early times when the half-light radius is within the core radius. After the linear approximation breaks down, we need to use the prediction in including the effect of the soliton. Importantly, we can see the analyses which excludes the soliton and only uses the quasi-particle approximation is inaccurate particularly at early times. This may be surprising when comparing the accuracy of this prediction for large half-light radii, but this result underscores the importance of altering our predictions when the quasi-particle approximation breaks down. 

We would also note the dependence of the heating rate on the specific halo model. For some objects the measurements of the half-light radius and velocity dispersion have large error bars. And so we would comment on how varying these parameters changes the predicted heating. We find that our results are most sensitive to measurements of the velocity dispersion, and the initial radius of the stellar distribution. The velocity dispersion measures how much dark matter is present to produce heating and is therefore intuitively very important to constraints on dark matter microphysics. Interestingly, this rate can affect both the heating rate near and far from the central soliton. In Figure \ref{fig:compareDispersion}, we compare two halos with different velocity dispersions. We can see this results in a different heating rate in both regimes. We would simply note the sensitivity of heating rates to measurements of velocity dispersion. Additionally, we have not simulated the regime in which the stellar half-light radius is small compared to the de Broglie wavelength and the stellar density is large compared to the dark matter density. However, this regime may be important (particularly given the choice of initial half-light radius in the simulations in \cite{Dalal2022}) and we treat it analytically in Appendix \ref{app:smallR}. In this appendix we see that the heating rate may depend on sensitively on the initial stellar density (and hence the initial radius of the stellar distribution).

In Appendix \ref{app:segueii} we use quasi-particle approximation, i.e., equation \eqref{eqn:dE_far}, to reproduce the results for the growth of the half-light radius in Figure 2 presented in \cite{Dalal2022}. This is simply a verification to demonstrate that our predictions reproduce existing results in the literature. 

\section{Conclusions} \label{sec:conclusion}

In this work we have simulated and provided analytic estimates of the heating of stellar particles due to the fluctuating gravitational potential in an ultralight dark matter halo. We have studied the heating due to the central soliton with simulations. 
We find that when a centrally located star cluster has a half-light radius comparable to or smaller than the soliton radius, the heating is not well described by the quasi-particle approximation. We provide instead a phenomenological formula that captures the heating rate observed in simulations.

We have studied the effect of stellar self-gravity with simulations, and provided an analytic method to approximate the impact on the growth of half-light radius.
We find that the stellar self-potential can slow the half-light radius growth rate when the stellar mass density is similar to or larger than the dark matter density. It's worth emphasizing that a stellar system that is dark matter dominated today could be star dominated in the past. 

We have simulated and studied the heating in tidally stripped halos. We find that the heating rate is substantially reduced if a significant fraction of the halo outside the soliton is stripped away.



More work will need to be done before we can assess the impact of these systematic effects on the existing heating constraints on ultralight dark matter.
First, the tidal stripping of ultralight dark matter subhalos should be studied in more detail. From earlier works, we already know that the tidal disruption of solitons is enhanced by quantum pressure \cite{Hui_2017,Du:2018qor}. The tidal disruption rate of subhalos, accounting for both the soliton and outer granules, should be better quantified. A computation in this direction was recently done by 
\cite{yang2025} for Fornax.
Second, the limit of a small stellar half-light radius compared to the de Broglie wavelength (or soliton radius) deserves a careful numerical study. It's a challenging problem because of the separation of scales involved. Our rough analytic estimate in Appendix \ref{app:smallR} suggests self-gravity and tidal field suppression could modify the half-light radius growth substantially. This is particularly relevant for the low end of the dark matter mass spectrum, with its associated long de Broglie wavelength. We hope to address these issues in the near future.

Let us close by emphasizing that the heating rate depends sensitively on the dark matter density inferred from dynamical measurements.
For systems with limited dynamical data, a wider range of heating rates is possible.
As an example, \cite{Segue1BH} recently pointed out that the Segue 1 dynamical data can be fitted by a model dominated by a $\sim 4 \times 10^5 {\,\rm M_\odot}$ black hole. 
If this is correct, the dark matter density in the system could be significantly lower,
leading to reduced heating. 

\appendix*
\renewcommand{\thesubsection}{\Alph{subsection}}
\section{}

\subsection{Plummer sphere initial conditions} \label{app:PlummerICs}
When determining a particle's position, we choose three random numbers, $X_1, X_2, X_3 \in (0,1)$. The first corresponds to $X_1 = M(r)/M_{\rm tot}$ which ensures that the density profile in equation (\ref{eq:Plummer_rho}) is obtained. To get the dimensionless radius, equation (\ref{eq:Mofr}) is inverted:
\begin{equation}
    a \equiv \frac{r}{R} = \left(X_1^{-2/3}-1\right)^{-1/2}.
\end{equation}
For a star on a spherical shell radius $a$, the angular coordinates are now obtained from the other two uniformly distributed variables $X_2$ and $X_3$. In cartesian coordinates this corresponds to
\begin{align}
    z&=a(1-2 X_2)\\
    x&= \sqrt{a^2-z^2} \cos{(2\pi X_3)}\\
    y&= \sqrt{a^2-z^2} \sin{(2\pi X_3)}.
\end{align}
To determine velocities, we need to ensure that they follow the isotropic distribution function
\begin{equation}
    f({\bf r},{\bf v})= \frac{24\sqrt{2}}{7\pi}\frac{R^2}{G^5 M^4} (-E)^{7/2}
\end{equation}
where $E = \Phi(r)+\frac{v^2}{2}$ is the energy of a star per unit mass. Probability function can in general be integrated to give the mass via
\begin{equation}
    dM = f({\bf r},{\bf v}) \, d^3{\bf r} \, d^3{\bf v},
\end{equation}
or for a fixed $r$ and isotropic velocities we can write
\begin{equation}
    \left. dM \right|_{r}=f(r,v)  4\pi v^2 dv,
\end{equation}
where $v$ is the speed of a star, $v= |{\bf v}|$. 
At a radius $r$, the escape velocity is
\begin{equation}\begin{split}
v_e(r) &= \sqrt{-2 \Phi(r)} \\
&= \sqrt{\frac{2 G M_{\rm tot}}{R}} \left( 1+a\right)^{-1/4}
\end{split}\end{equation}
so we can write with the help of dimensionless variable for the speeds, $q = v/v_e$: 
\begin{equation}\begin{split}
    \left. dM \right|_{r} &= \alpha \: q^2 (1 - q^2)^{7/2}\\
    &=\alpha \:g(q),
\end{split}\end{equation}
where $\alpha$ contains numerical factors and dimensional factors $R$, $M_{*}$, and $v_e(r)$. Function $g(q)$ is defined on the interval $q \in (0,1)$ and never exceeds $\sim0.1$. To satisfy the above distribution, we follow the von Neumann rejection technique. Two random numbers are chosen such that $X_4, X_5 \in (0,1)$. If the pair satisfies $g(X_4) > 0.1 X_5$, we keep $X_4$, otherwise we continue selecting random pairs until the condition is met. We then assign $q=X_4$ and for the speed, $v = v_e q$. Two more random numbers $X_6, X_7 \in (0,1)$ then allow us to calculate the velocity:
\begin{align}
    v_z&=v(1-2 X_6)\\
    v_x&= \sqrt{v^2-v_z^2} \cos{(2\pi X_7)}\\
    v_y&= \sqrt{v^2-v_z^2} \sin{(2\pi X_7)}.
\end{align}
Finally, we make sure that positions and velocities have the right units and that the virial theorem is satisfied.
Plummer's model is increasingly more correct with the increasing number of stars. But for system with smaller number of stars, randomness of a realisation can in principle introduce deviations from the virial theorem,  $2K + W = 0$. A possible solution is to rescale the velocities and ensure the cluster is viralised; velocities of all stars are simply multipiled by the factor of $\sqrt{|W|/2K}$.
To obtain the total potential and kinetic energies for a collection of stars, we can assume that the gravitational potential $\Phi$ is nearly spherically symmetric. With this, the potential and kinetic energies are
\begin{equation}
W = \text{trace}\left( -\int \rho(\bar{x})x_j \frac{\partial \Phi}{\partial x_k} d^3x \right) = -\sum_{i=\text{stars}} m_i r_i \frac{\partial \Phi}{\partial r}
\end{equation}
and 
\begin{equation}
K = \frac{1}{2} \sum_{i=\text{stars}} m_i v_i^2.
\end{equation}

\subsection{Reproducing existing results} \label{app:segueii}

\begin{figure}[!ht]
	\includegraphics[width = .47\textwidth]{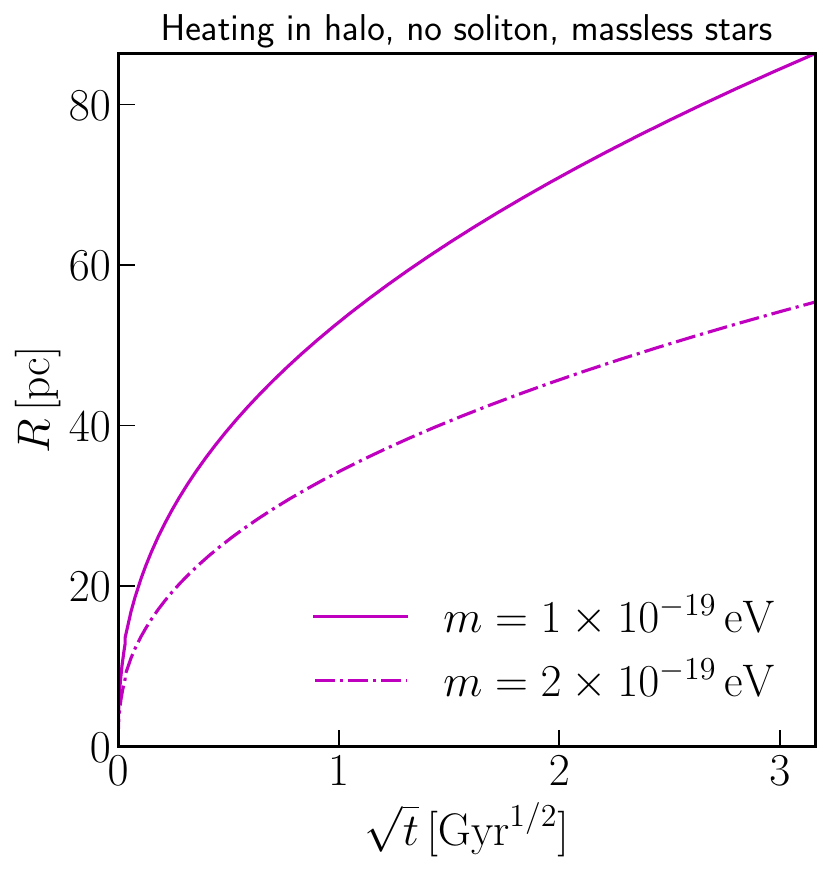}
	\caption{ Analytic predictions for the growth of the half-light radius in a system simulated by \cite{Dalal2022}. For each line, the halo parameters are chosen such that the orbital velocity at radius $R=40 \, \mathrm{pc}$ is $3 \, \mathrm{km/s}$ and the concentration is $c=20$, though our results are not sensitive to small changes in these choices. The initial half-light radius is chosen to be $1 \, \mathrm{pc}$.}
	\label{fig:segueIIPredict}
\end{figure}

In this Appendix, we present a test of the method described in Section
\ref{sec:discussion}. Using our method, we compute the prediction of the half-light-radius growth for a system simulated in \cite{Dalal2022}. We make the same assumptions as \cite{Dalal2022}: massless stars (i.e. stars as purely test particles)
and no soliton. Our results, shown in Figure \ref{fig:segueIIPredict}, match those
in Figure 2 of \cite{Dalal2022}. 
Note that strictly speaking, when the initial radius is as small as $1$ pc,
the self-gravity of the stars cannot be ignored. This is explored
in the following Appendix.

\subsection{Accounting for tidal field suppression and stellar self-gravity: application to Segue 1} \label{app:smallR}

There are two effects we wish to explore in this Appendix.
First, if the stellar cluster of interest is initially compact enough that the
stellar self-gravity is important, the stars can no longer be
treated as simply test particles. Under the influence of its own gravity,
we expect the stellar cluster's expansion by heating to be suppressed.
Second, if the stellar cluster's size $R$ is small compared to
the de Broglie wavelength, the heating should be further suppressed.
This is because as far as the internal dynamics of the stellar cluster is
concerned, it's the differential accerleration (due to the tidal field
exerted by the dark matter substructure) that counts.
This tidal field suppression works in the same way as in standard
tidal disruption computations: the smaller $R$ is, the more strongly
suppressed is the differential acceleration.
This effect should be distinguished from 
another effect discussed earlier, that due to the tidal stripping of halos.


Let us incorporate the effect of tidal field suppression first.
Consider the differential acceleration on two stars separated by a distance $R$ incident on a granule of density $\rho$ and radius $\lambda_\mathrm{dB}$. The differential acceleration is given by
\begin{equation}
\label{eq1}
\delta a =  {G [(4\pi/3) \rho \lambda_{\rm dB}^3] \over \lambda_{\rm dB}^2}  {R \over \lambda_{\rm dB}}
\end{equation}
The factor of $R/\lambda_{\rm dB}$ captures the effect of tidal field suppression, assuming $R \ll \lambda_{\rm dB}$. If $R > \lambda_{\rm dB}$, then the forces on the stars due to passing granules should be uncorrelated and this factor should just be replaced by $1$, i.e. $R/\lambda_{\rm dB} \rightarrow 1$.

The differential velocity change the stars would receive over the lifetime of the granule is the differential acceleration multiplied by the de Broglie time $\tau_\mathrm{dB} = \lambda_\mathrm{dB} / \sigma$. 
Adding multiple kicks in a random walk fashion, the net root-mean-squared differential velocity change over the time $\Delta t$ is
\begin{equation}
  \Delta v \sim {G [(4\pi/3) \rho \lambda_{\rm dB}^3] \over \lambda_{\rm dB}^2}  {R \over \lambda_{\rm dB}}
  {\lambda_{\rm dB} \over \sigma} \sqrt{\Delta t \over \lambda_{\rm dB}/\sigma} \, .
\end{equation}
Or equivalently:
\begin{equation}
\label{granulefulltidal}
\Delta v^2 \sim {G^2 [(4\pi/3) \rho]^2 R^2 \lambda_{\rm dB} \over \sigma} \Delta t \, .\end{equation}
This expression can be viewed as a master formula which can be applied in different contexts. It accounts for tidal field suppression, when
$R \ll \lambda_{\rm dB}$. 
If $R \, \gsim \, \lambda_{\rm dB}$, there is no such suppression, in which case
\begin{equation}
\label{granulenotidal}
\Delta v^2 \sim {G^2 [(4\pi/3) \rho]^2 \lambda_{\rm dB}^4 \over \sigma} \Delta t \, \, \end{equation}
which reproduces \eqref{eqn:deltaVar}. 

Let us apply the master formula to a stellar cluster inside a soliton, with the understanding that the heating by the random walk and oscillation of the soliton
can be approximately captured by the granule-based derivation that leads to \eqref{granulefulltidal} \cite{Li:2020ryg}.
Recall that for a soliton embedded in a halo, the soliton radius roughly matches the de Broglie wavelength $\lambda_{\rm dB}$ of the halo, and
$G[(4\pi/3) \rho \lambda_{\rm dB}^3]/\lambda_{\rm dB}\sim \sigma^2$
with $\sigma$ being the velocity dispersion of the halo.
Therefore, \eqref{granulefulltidal}  can be written as
\begin{equation}
\label{fulltidal}
\Delta v^2 \sim { G [(4\pi/3) \rho R^3] \over R} {\Delta t \over \lambda_{\rm dB}/\sigma} \, .
\end{equation}
This describes the heating of a stellar cluster inside a soliton,
with $R$ much less than the soliton radius.

A parenthetical remark:
If we had applied the same soliton scaling relation
to \eqref{granulenotidal} (which has the tidal
field suppression turned off), we would have obtained
\eqref{fulltidal}, but with an extra multiplicative factor
of $\lambda_{\rm dB}^2/R^2$.
On the other hand, the phenomenological fit of equation \eqref{eqn:dE_nearCore}
is the same as (up to a numerical factor) \eqref{fulltidal} times
a factor of $\lambda_{\rm dB}/R$ instead. Thus, the phenomenological
fit appears to have neither full-on nor full-off tidal field suppresion, but somewhere in between.
It describes the heating when $R$ is not much smaller than $\lambda_{\rm dB}$.

Let us apply \eqref{fulltidal} to the heating of a self-gravitationally bound stellar cluster, i.e. one in which the mass is dominated by the stars themselves.
Let its mass be $M_*$, and radius be $R$. The total energy (per unit mass)
of the self-bound stellar cluster is about $\sim - GM_*/R$, ignoring factor
of order unity. Now, equation \eqref{fulltidal} tells us the amount of energy (per unit mass) being pumped into the system. When the amount of energy
roughly matches $GM_*/R$, the system would become unbound.
Let us call the time for disruption $\Delta t_{\rm disrupt}$. We can thus estimate it from:
\begin{equation}
{GM_* \over R} \sim { G [(4\pi/3) \rho R^3] \over R} {\Delta t_{\rm disrupt} \over \lambda_{\rm dB}/\sigma} \, ,
\end{equation}
from which we deduce
\begin{equation}
\label{Dtdisrupt}
\Delta t_{\rm disrupt} \sim {M_* \over [(4\pi/3) \rho R_i^3]} {\lambda_{\rm dB} \over \sigma} \, .
\end{equation}
We have included the subscript $i$ in $R_i$, to emphasize this is the initial
radius of the stellar cluster (before it starts expanding due to heating).
So, the time to disruption is the de Broglie time multiplied by the ratio of
stellar mass to dark matter mass enclosed within $R_i$.

Let's plug in some numbers motivated by Segue 1.
Its observed density today, which is dominated by dark matter, is
around $\rho \sim 1 {\rm \, M_\odot} / {\rm pc}^3$, with a stellar mass of about
$M_* \sim 10^3 {\,\rm \, M_\odot}$.
\footnote{The precise numbers depend on modeling assumptions \cite{Geha:2008zr}. According to one way of fitting the
  data, the mass enclosed within $50$ pc is
  $8.7 \times 10^5 {\,\rm M_\odot}$, which gives $\rho \sim 1.67 {\,\rm M_\odot / pc^3}$. The stellar mass of $\sim 10^3 {\,\rm M_\odot}$ is estimated based on the absolute magnitude of ${M_V} \sim -1.5$ (with substantial errorbar of $+0.6$ and $-0.8$), and a stellar-mass-to-light ratio of $3 {\,\rm M_\odot / L_\odot}$. See \cite{Geha:2008zr} for details.}
Suppose the initial radius of the stellar cluster is $R_i \sim 1$ pc,
which corresponds to an initial stellar mass density of
$\rho_* {}_i \sim 240 {\,\rm M_\odot / pc^3}$:
\footnote{See \cite{Strader2013} for an example of high, in fact higher, stellar density.}
thus initially,
the self-gravity of the stars dominates over the gravity from dark matter.
We will approximate the dark matter density $\rho$ as roughly constant
even as the stellar cluster gets heated and expands.
We will also assume the wave dark matter mass is low enough that
$\lambda_{\rm dB}$, or roughly the soliton radius, is much larger than $R_i$,
\footnote{\label{solitonscales}
  It can be checked that for a soliton with core density $\rho$,
  the corresponding core radius is $\sim 180 {\,\rm pc\,}
  (10^{-21} {\,\rm eV}/m)^{1/2} ({\rm M_\odot/pc^3} / \rho)^{1/4}$,
  and the soliton mass is $\sim 2 \times 10^7 {\,\rm M_\odot}
  (10^{-21} {\,\rm eV}/m)^{3/2} (\rho / {\rm M_\odot/pc^3})^{1/4}$ \cite{Hui_2017}. 
}
such that tidal field suppression is important.
The last piece of information we need for \eqref{Dtdisrupt} is
the de Brogile time $\lambda_{\rm dB}/\sigma$, which can
be estimated using the dynamical time of the soliton, i.e.
$1/\sqrt{G\rho} = 1.5 \times 10^7 {\,\rm yr.} ({\,\rm M_\odot/pc^3}/\rho)^{1/2}$.
Therefore,
\begin{equation}
\label{Ddisrupt2}
\Delta t_{\rm disrupt} \sim 4 \times 10^9 {\,\rm yr.}
\left( {\rho_* {}_i \over 240 {\,\rm M_\odot / pc^{3}}} \right)
\left( {\,\rm M_\odot /pc^{3}} \over \rho \right)^{3/2} \, .
\end{equation}
This estimate is largely independent of the wave dark matter mass $m$.
The only assumption is that the initial stellar cluster radius is small
compared to the soliton radius, which holds for a wide range
of dark matter masses (see footnote \ref{solitonscales}).
It also assumes $\rho_* {}_i > \rho$, such that stellar self-gravity
dominates initially.

That this {\it very rough} estimate is a sizable fraction of the age
of the universe is interesting. This means one could envision a scenario
of how the current Segue 1 configuration comes about:
it originates as a self-bound stellar cluster inside a soliton, which spends a long
time more or less unaffected by the persistent heating---
until recently. Once it gets disrupted, the estimates in
Section \ref{sec:insideCore} suggest the radius would grow rather quickly,
and we happen to be catching Segue 1 at this growing phase.

A drawback of this scenario is that some degree of fine-tuning is involved: we have to catch Segue 1 just recently reaching the current size. 
When a large sample of similar objects is available, one could start asking questions about the statistical plausibility of this scenario. Perhaps a more important drawback is this:
the ratio of the stellar dynamical time to the de Broglie time is roughly
$(\rho / \rho_* {}_i)^{1/2} \sim 0.06 (240 \rho/\rho_* {}_i)^{1/2}$.
This means the gravitational potential fluctuates on a time scale
long compared to the dynamical time of the stars.
In that case, one would think the heating rate should be suppressed.
A reliable estimate would require numerical simulations, which we hope to carry out in the near future.

\bibliography{BIB}

\end{document}